\documentclass[12pt,vi,twoside]{mitthesis}
\usepackage[sort,numbers]{natbib}
\usepackage{graphicx}
\usepackage{epstopdf}
\usepackage{amsmath}
\usepackage{amssymb}
\usepackage{booktabs}
\usepackage{dcolumn} 
\usepackage{url}
\usepackage{caption}
\usepackage{subcaption}
\usepackage{aas_macros}
\usepackage[colorlinks=true]{hyperref}  
\hypersetup{linkcolor=black, citecolor=black, filecolor=black, urlcolor=black}

\newcommand{\ra}[1]{\renewcommand{\arraystretch}{#1}} 
\newcolumntype{.}{D{.}{.}{-1}} 

\newcommand{\Sersic}{S\'{e}rsic }

\begin{document}

%
%
%
%
%
%
%
%
%
%
%

\title{ \uppercase{Optical Detection and Analysis of Pictor A's Jet}}

\author{ \uppercase{Eric S. Gentry}}
\department{ \uppercase{Department of Physics}}

\degree{ \uppercase{Bachelor of Science}}

\degreemonth{June}
\degreeyear{2014}
\thesisdate{May 9, 2014}


\supervisor{Doctor Herman L. Marshall}{Principal Research Scientist}

\chairman{Professor Nergis Mavalvala}{Senior Thesis Coordinator, Department of Physics}

\maketitle



\cleardoublepage
\setcounter{savepage}{\thepage}
\begin{abstractpage}

%
%
%
New images from the Hubble Space Telescope of the FRII radio galaxy Pictor A reveal a number of jet knot candidates which coincide with previously detected radio and x-ray knots.  Previous observations in x-ray and radio bands show the entire jet to be 1.9\arcmin{} long, with interesting variability, but an optical component was previously unknown.  The discovered optical component is faint, and knot candidates must be teased out from a bright host galaxy.  Using three broadband filters, we extract knot fluxes and upper-bounds on the flux for multiple knot candidates at wavelengths of 1600nm, 814nm and 475nm.  We find that the data suggest that localized particle re-accleration events followed by synchrotron emission could explain the observed knot candidates, but those electrons could not supply enough x-ray flux to match prior observations. Our data provide key evidence suggesting a second, higher energy electron population which was previously hypothesized, but could not be confirmed.
\end{abstractpage}


\cleardoublepage

\section*{Acknowledgments}

First and foremost, I would like to thank my thesis adviser, Herman Marshall, for his help and guidance, and for the opportunity to work on this project. I also extend my thanks to the High Energy Astrophysics group at MIT, with whom I have worked for the past year.

For the writing process in particular, I would like to thank Jamie Teherani, the Graduate Resident Tutor of my undergraduate living group, for helping provide a guiding framework with which to approach the writing of this thesis.

For developing my interest in astronomy, I am indebted to professors Paul Schechter and Rob Simcoe, my academic adviser. They helped me find my interest in astronomy, which resulted in this thesis.

Finally, I would like to thank my parents, Abbie and Stuart Gentry, for their support and guidance throughout my studies and the writing of this thesis.

(Support for this work was provided in part by NASA through grant GO-12261.01-A from the Space Telescope Science Institute, which is operated by the Association of Universities for Research in Astronomy, Incorporated, under NASA contract NAS5-26555.)


\pagestyle{plain}

\tableofcontents
\newpage
\listoffigures
\newpage
\listoftables


\chapter{Introduction}

Some of the most luminous objects of the universe are active galactic nuclei (\emph{AGN}), systems in which the central supermassive blackhole of a galaxy is accreting nearby material, releasing large amounts of energy in the process.  In many cases, the luminosity from this central region of a galaxy can exceed the luminosity of the rest of the galaxy combined.  This luminosity plays an important role both inside and outside the host galaxy:  Within a galaxy, an AGN can play an important role in slowing stellar formation \cite{Olsen:2012me}, and beyond the galaxy itself, AGN also play an important role in bringing about cosmic re-ionization of the intergalactic medium (\emph{IGM}) through photo-ionization.  In order to understand the impact of AGN, we must understand what we are observing, and that understanding comes through the study of radiative processes within high energy plasmas.

AGN are not found within every galaxy -- most nearby galaxies, including the Milky Way, currently have dormant supermassive blackholes.  It is believed though that most supermassive blackholes, at some point in their lives, become active (astronomers will be watching this year as the gas cloud \emph{G2} approaches our own galactic center, but that accretion event will be much smaller than the accretion onto typical AGN \cite{Gillessen12}).  Surveys suggest that AGN were more common at earlier times in the universe, but this poses a challenge to their study: most of the visible AGN are distant (with redshifts, \emph{z}, peaked at $z\approx1$ \cite{Ueda2014}), and are thus fainter and more difficult to resolve.  We would like to be able to peer into the structure of AGN, but that is difficult for all but the closest sources.

Pictor A is a particularly interesting galaxy because it is both nearby (redshift, $z$, of .035), and features from its AGN are larger than similar sources.  From the AGN core of Pictor A comes a jet of relativistically moving particles, which is about 1 kiloparsec (\emph{kpc}) wide, and extends hundreds of kiloparsecs.  For comparison, the nearby galaxy M87 ($z=.004$), has a well-studied jet which is only parsecs wide, and extends only a few kiloparsecs.  The combination of Pictor A's size and proximity means we are able to resolve multiple layers of substructure, probing a regime of high energy processes which are not well understood.

In this thesis we will look at recent observations of Pictor A, and make inferences about the radiative processes which are present.

This chapter will look at the background of AGN studies in greater depth, and examine how past research could inform our understanding of Pictor A. In turn, we will also discuss Pictor A, and how it could fill in some of the gaps in our understanding of AGN.

In particular, since AGN often have significant spectral, spatial, and temporal features, this chapter will discuss previous observations of Pictor A in Section \ref{intro:PictorA}.  This both provides motivation for the current work, and will factor directly into findings discussed in later chapters.  Section \ref{intro:outline} will then give an overview for the research which constitutes the body of this work.

Chapter \ref{theory} treats the subject broadly, giving an overview of the theory guiding this inquiry.

Chapter \ref{methodology} discusses the methodology behind the data collection, processing and analysis, as well as what systemic biases and uncertainties are introduced in our data.

Chapter \ref{results} discusses the observational and reduced results.

Chapter \ref{significance} will consider those reduced parameters in the context of previous observations and physical theories of underlying mechanisms.  Discussion of future analyses and recommended observations can be found in Appendix \ref{appendix:future}.

\section{Previous Observations of Pictor A}
\label{intro:PictorA}
Pictor A first attracted attention with its incredibly bright radio emission.  Early radio surveys identified it as one of the most luminous radio sources in the sky \cite{Robertson73}.  In 1987 it was found that one of its two bright radio lobes was spatially coincident with observed optical synchrotron emission \cite{Roeser87}, and one of the earliest x-ray surveys, the Einstein Imaging Proportional Counter survey (\emph{Einstein IPC}), also partially detected x-ray emission from this lobe \cite{Roeser87}.

\begin{figure}[tb]
	\begin{center}
		\includegraphics[width=.7\linewidth]{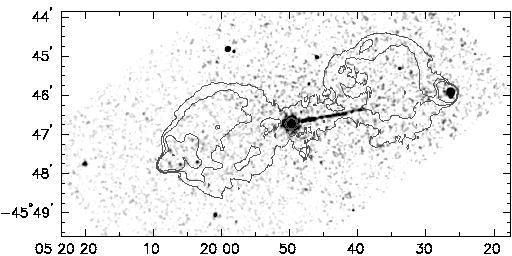}
	\end{center}
	\caption{Taken from Wilson et al. \cite{Wilson01}.  X-ray (Chandra) image overlay \cite{Wilson01}, on top of radio (VLA 20cm) contours \cite{Perley97}.  Radio lobes can be seen on either side, but the x-ray jet has only been detected on one side of Pictor A, possibly due to relativistic beaming.}
	\label{fig:XrayRadioMap}
\end{figure}

Evidence for a jet first came through radio observations, but was most conclusively shown in x-ray images.  The radio data came from the Very Large Array (\emph{VLA}), discovering a jet faintly visible extending to the western radio lobe \cite{Perley97}.  While faint in the radio, this jet was found to be surprisingly bright in Chandra x-ray observations \cite{Wilson01} (image shown in Fig.\ref{fig:XrayRadioMap}).  Follow-up Chandra observations revealed that a particular structure in the jet, a \emph{knot}, appeared to flare and fade on the timescales of years, whereas theory would have na\"{\i}vely predicted a lifetime on the order of thousands of years \cite{Marshall10}.  To resolve this, Marshall et al. suggested the flare occurred within an unresolved region within the broader jet (cf. flares observed by Godfrey et al. and Chang et al. \cite{Godfrey09, Chang10}).  In order to test this hypothesis, better angular resolution and more spectral information was required.

In 2010, Marshall et al. used the Hubble Space Telescope (\emph{HST}) to obtain better spatial resolution and additional spectral information on the nature of the jet.  The imaging resolution (around $.05 - .1$ arcseconds, \arcsec) is an improvement over Chandra's $.5$\arcsec{} resolution, and the chosen filters (with bandpasses centered on wavelengths $1.6\mu$m, $814$nm, $475$nm) were expected to be near a break in the synchrotron spectrum.

In our HST observations, we discovered unresolved substructure, and the photometry revealed a sharp synchrotron cutoff.  The significance of these findings will be discussed later.

\section{Outline of this Research}
\label{intro:outline}

\begin{figure}[tbp]
\centering
	\begin{subfigure}{.3\linewidth}
		\includegraphics[width=\textwidth]{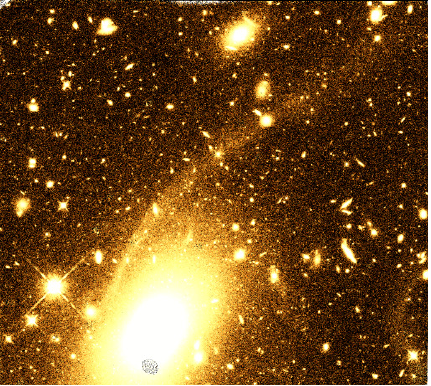}
		\label{fig:samplegalaxysubtraction:galaxy}
		\caption{Raw image}
	\end{subfigure}
	\quad 
	\begin{subfigure}{.3\linewidth}
		\includegraphics[width=\textwidth]{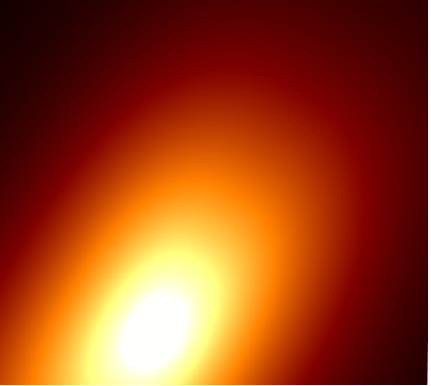}
		\label{fig:samplegalaxysubtraction:model}
		\caption{Model}
	\end{subfigure}
	\quad
	\begin{subfigure}{.3\linewidth}
		\includegraphics[width=\textwidth]{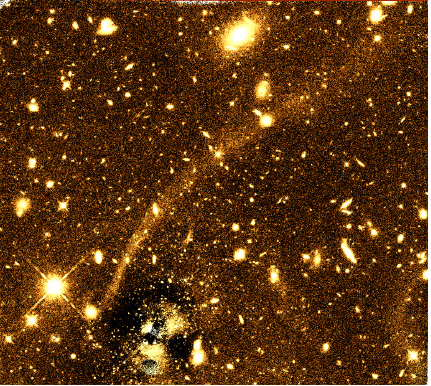}
		\label{fig:samplegalaxysubtraction:resid}
		\caption{Residual}
	\end{subfigure}

	\caption{Optical images demonstrating galaxy subtraction. Images are false-color (seen through one band) with a color map and a logarithmic stretch function matching the style of other galaxy-subtracted images (see \cite{GALFITNearby}).}
	\label{fig:GalaxySubtraction}
\end{figure}

This research focuses on 3 HST images of Pictor A, in optical bands ranging from infrared (\emph{IR}) to ultraviolet (\emph{UV}).  From these images, we want information about emission coming from relatively atypical jet knots, but those knots are largely drowned out by the large surface brightness of the otherwise typical host galaxy.  In order to better identify knot emission, it is then necessary to subtract a model for a ``typical'' elliptical galaxy's surface brightness (see Figure \ref{fig:GalaxySubtraction}), in order to see the atypical features that remain.

\begin{figure}[tbp]
	\begin{center}
		\includegraphics[width=.8\linewidth]{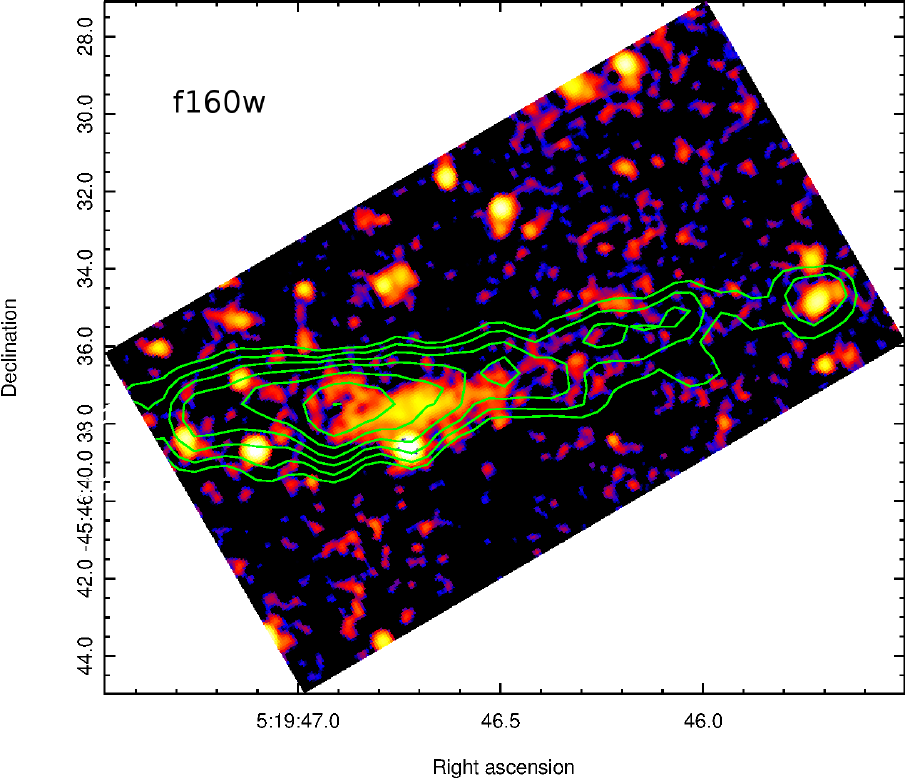}
	\end{center}
	\caption{Sample knot 32\arcsec{} from the AGN. Optical image with x-ray contours overlaid.  This image is also a false-color image, with a logarithmic  stretch function.  All future images will also be false-color with logarithmic stretch functions, except when otherwise noted.}
	\label{fig:sample_knot_f160w_32arcsec_intro}
\end{figure}

Using that \emph{galaxy-subtracted} image, we identified features which appeared spatially coincident with observed x-ray and radio features (see knot example, Figure \ref{fig:sample_knot_f160w_32arcsec_intro})  Using these knots, we determined the fluxes through each band.  This photometric information was then used to test analytic and numeric models of relativistic synchrotron radiation, synchrotron self-absorption (\emph{SSA}), synchrotron self-Compton radiation (\emph{SSC}), and inverse Compton scattering of the Cosmic Microwave Background (\emph{IC-CMB}).

\chapter{Theory}
\label{theory}
In this chapter, I will summarize (and when necessary, derive) the key theoretical frameworks guiding this analysis.  Chapter \ref{methodology} will discuss the implementation of the theory discussed in this chapter.

\section{Galaxy Subtraction}
In order to understand what was unusual about this galaxy, we first needed to be able to discriminate between what was normal and what was abnormal.  In order to do that, we fit standard galaxy morphologies to our data, and analyzed the residual.  Key to this process was understanding common galaxy morphologies, and how those morphologies might relate to our particular galaxy.

\subsection{Galaxy Morphologies}
\label{theory:morphology}
In the 1920s, Hubble noticed that galaxies can often be grouped into two main categories: ellipticals and disks, in a classification now known as the \emph{Hubble Sequence} \cite{HubbleSequence}.  The specific surface brightness (as a function of radius), $\Sigma_\nu(r)$, of these respective categories are commonly fit by two profiles:

\begin{equation}
	\Sigma_\nu(r) \sim
	\begin{cases}
		\exp\left(-\left[\frac{r}{r_0}\right] \right) & \mathrm{disk} \\
		\exp\left(-\left[\frac{r}{r_0}\right]^{1/4} \right) & \mathrm{elliptical} \\
	\end{cases}
	\label{eq:SurfaceBrightnessProfiles}
\end{equation}

In 1963 \Sersic generalized these profiles, and developed the \emph{\Sersic index} \cite{SersicIndex}, which now is used to quantitatively describe galactic morphology.  The surface brightness profile for an arbitrary \Sersic index, $n$, is:

\begin{equation}
	\Sigma_\nu(r) \sim \exp\left(-\left[\frac{r}{r_0}\right]^{1/n} \right)
	\label{eq:SersicProfile}
\end{equation}

The versatility of the \Sersic profile results in its usefulness as a generic galaxy morphology model.  In particular, when looking at atypical aspects of a galaxy, a \Sersic profile can be subtracted to leave only the unusual aspects as residuals \cite{GALFIT}. Multiple \Sersic profiles (with different \Sersic indices, magnitudes, and scale radii, $r_0$) can then be combined to provide a useful, if incomplete basis set of models for describing a galaxy.

\section{Radiative Transfer}
\label{theory:radiativetransfer}
The following sections will summarize the field of high energy radiative transfer of non-thermal emission, as it applies to relativistic jets.  In some cases, a more thorough handling is more appropriately left to review articles and books; in those cases, I refer the reader to appropriate sources.  What is presented is a summary of key points, and derivations for aspects not typically discussed.  In general, I will try to present equations in dimensionless form (following the style of \cite{Gould1979}), but when that is not possible, it is safe to assume Gaussian cgs units (e.g. spectral flux density, $S_\nu$, has units of erg s$^{-1}$ cm$^{-2}$  Hz$^{-1}$)

Before beginning, we assume a homogeneous plasma of charged particles of energies $E = \gamma m c^2$ (for a distribution of $\gamma$), moving through a constant magnetic field, $B$, with a pitch angle, $\alpha_{pitch}$, between their momenta and the local magnetic field. (We will refer to energies and Lorentz factors, $\gamma$, interchangeably, as they are trivially related by $E = \gamma m c^2$; it is easier to think of energy, but easier to express $\gamma$, which is independent of your choice of units).

We can assume electrons (and positrons) are responsible for a majority of the observed radiation, given their low mass (allowing for high accelerations, required to produce non-thermal radiation).  For a number density of electrons, $n_e$, their distribution of energies is typically chosen to be a power law distribution with electron energy index $p$, and with low and high energy cutoffs imposed:

\begin{equation}
	n_\gamma \equiv \frac{d n_e}{d \gamma} =
	\begin{cases}
		\kappa_e \gamma^{-p} & \gamma_\mathrm{min} < \gamma < \gamma_\mathrm{max} \\
		0  & \mathrm{else}
	\end{cases}
	\label{eq:ElectronDistributionWithCutoffs}
\end{equation}

We may now begin to look at the predictions made by the hypothesized emission mechanisms: synchrotron radiation and inverse Compton up-scattering of background photons.

\subsection{Synchrotron Radiation}
\label{theory:synchrotron}
A free charge moving non-relativistically through a magnetic field will experience a Lorentz force:

\begin{equation}
	\mathbf{F} = q(\mathbf{E} + \frac{\mathbf{v}}{c} \times \mathbf{B})
	\label{eq:LorentzForce}
\end{equation}

In a constant magnetic field, the general solution for a freely moving charge governed by Equation \ref{eq:LorentzForce} is a helix, with a pitch angle, $\alpha_{pitch}$, between the magnetic field and the particle's velocity. It can then be shown that circular motion is described by $\alpha_{pitch} = \frac{\pi}{2}$ and linear motion is described by $\alpha_{pitch} = 0$.  Under these conditions, a charge $q$ of mass $m$ will have a gyro-frequency (and thus an emission frequency) of:

\begin{equation}
	\nu_g = \frac{q B}{2 \pi m c}
	\label{eq:gryofrequency}
\end{equation}

When treated relativistically (with $\gamma = \frac{E}{m c^2}$), we find the more general fundamental frequency:

\begin{equation}
	\nu_r = \gamma \nu_g \sin\left( \alpha_{pitch}\right)
	\label{eq:synchrotronfrequency}
\end{equation}
with higher order harmonics (derived in \cite{LongairBook2ed}) leading to a maximal emission peak for a relativistic electron at a frequency:
\begin{equation}
	\nu_{max} = .42 \gamma^2 \nu_g \sin\left( \alpha_{pitch}\right)
	\label{eq:synchrotronfrequencymax}
\end{equation}

In astrophysical contexts though, we are often interested in an ensemble of electrons of different energies and pitch angles.  The key results are well known (see \cite{LongairBook2ed,RybickiLightmanBook,BlumenthalGould1970}), but best summarized in \cite{Gould1979} under the assumptions of an isotropic spherically symmetric ensemble of electrons obeying the energy distribution in Equation \ref{eq:ElectronDistributionWithCutoffs}. (While Gould allows for a inhomogeneous source in \cite{Gould1979}, we will restrict ourselves to a homogeneous source for simplicity).  Under those conditions, we expect an (optically thin) flux density spectrum:
\begin{equation}
	S_\nu = \frac{2 \pi \hbar}{3}\nu \rho_\nu \frac{R^3}{d_L^2}
	\label{eq:gouldfluxspectrum}
\end{equation}
for a source of radius, $R$, at a luminosity distance, $d_L$, and which has a specific volumetric production rate of photons:
\begin{equation}
	\rho_\nu = 4 \pi \left(\frac{3}{2}\right)^\frac{p-1}{2} a\!\left(p\right) \alpha_{FS} \kappa_e \left( \frac{\nu_g}{\nu}\right)^\frac{p+1}{2}
	\label{eq:gouldrhonu}
\end{equation}
where $\alpha_{FS} \equiv e^2 / h c$ is the fine structure constant, and $a(p)$ is given in \cite{Ginzburg65} to be:
\begin{equation}
	a(p) = \frac{2^\frac{p -1}{2}}{8 \left(p+1\right)} \sqrt{\frac{3}{\pi}} \frac{\Gamma\left( \frac{3p -1}{12} \right) \Gamma\left( \frac{3p +19}{12} \right) \Gamma\left( \frac{p +5}{4} \right)}{\Gamma\left( \frac{p+7}{4} \right)}
	\label{eq:BlumenthalGoulda}
\end{equation}
leading to a power law spectrum:
\begin{equation}
	S_\nu \sim \nu^{-\frac{p-1}{2}} = \nu^{- \alpha_{sync}}
	\label{eq:synchrotronunabsorbedpowerlaw}
\end{equation}
which defines the synchrotron spectral index:
\begin{equation}
	\alpha_{sync} \equiv \frac{p-1}{2}
	\label{eq:synchrotronindex}
\end{equation}

It is important to note the range of validity of Equation \ref{eq:synchrotronunabsorbedpowerlaw}. The derivation assumed an unbounded electron power law (which is non-normalizable) -- that approximation is good as long as we are far from significant contributions from the bounds of our electron power law: $\gamma_{min}^2 \nu_g \ll \nu \ll \gamma_{max}^2 \nu_g$.  At high frequencies, Equation \ref{eq:synchrotronfrequencymax} predicts a cutoff at:
\begin{equation}
	\nu_{max} \approx \gamma_{max}^2 \nu_g
	\label{eq:synchrotronnumaxgammamax}
\end{equation}
A similar cutoff exists at low frequencies, but more commonly we find that the low frequency departure from Equation \ref{eq:synchrotronunabsorbedpowerlaw} is due to the source becoming optically thick, a possibility for which we did not account.

\subsubsection{Synchrotron Self-Absorption}
\label{theory:synchrotronselfabsorption}

Properly accounting for optical depth (see \cite{LongairBook3ed} for a detailed derivation) we expect a specific optical depth, $\tau_\nu$, after passing through a length, $\ell$, of the form \cite{Gould1979}:
\begin{equation}
	\tau_\nu = c(p)r_e^2 \kappa_e \frac{\nu_e}{\nu} \left( \frac{\nu_g}{\nu} \right)^\frac{p+2}{2} \ell
	\label{eq:synchrotronopticaldepth}
\end{equation}
where $r_e = e^2 / m_e c^2$ is the standard classical electron radius $\nu_e = c / r_e$ is a less standard definition used to simplify the equation and its units, and $c(p)$ is given by \cite{Gould1979} to be:
\begin{equation}
	c = \frac{3^{\frac{p+1}{2}}}{8} \sqrt{\pi} \frac{ \Gamma\left(\frac{p+6}{4}\right)\Gamma\left(\frac{3p+2}{12}\right)\Gamma\left(\frac{3p+22}{12}\right)}{\Gamma\left(\frac{p+8}{4}\right)}
	\label{eq:Gouldc}
\end{equation}

The process of a source both emitting, then reabsorbing synchrotron radiation is called \emph{synchrotron self-absorption} (\emph{SSA}).  Equation \ref{eq:synchrotronopticaldepth} predicts that this is more likely to dominate at low frequencies.  This defines a frequency, $\nu_{SSA}$ (such that $\tau_{\nu_{SSA}} = 1$), determining the transition between the low-frequency optically thick regime ($\tau_\nu > 1$) and the high-frequency optically thin regime (following Equation \ref{eq:synchrotronunabsorbedpowerlaw}).  This transition frequency can be found using Equation \ref{eq:synchrotronopticaldepth}:
\begin{equation}
	\nu_{SSA} = \left( c(p) r_e^2 \kappa_e \nu_e \nu_g^\frac{p+2}{2} R \right)^\frac{2}{p+4}
	\label{eq:nuSSA}
\end{equation}

For a more complete discussion of the effect of a specific optical depth, $\tau_\nu$, see \cite{LongairBook3ed}, but ultimately it modifies the spectrum given in Equation \ref{eq:synchrotronunabsorbedpowerlaw} to one that has the limiting behaviors:

\begin{equation}
	S_\nu \sim
	\begin{cases}
		\nu^{5/2} &  \nu_g \gamma_{min}^2 \ll \nu \ll \nu_\mathrm{SSA} \\
		\nu^{-\frac{p + 1}{2}}  & \nu_{SSA} \phantom{\gamma} \ll \nu \ll \nu_g \gamma_{max}^2 \\
	\end{cases}
	\label{eq:SynchrotronSpectrumFull}
\end{equation}

\begin{figure}[htbp]
	\begin{center}
		\includegraphics[width=.7\linewidth]{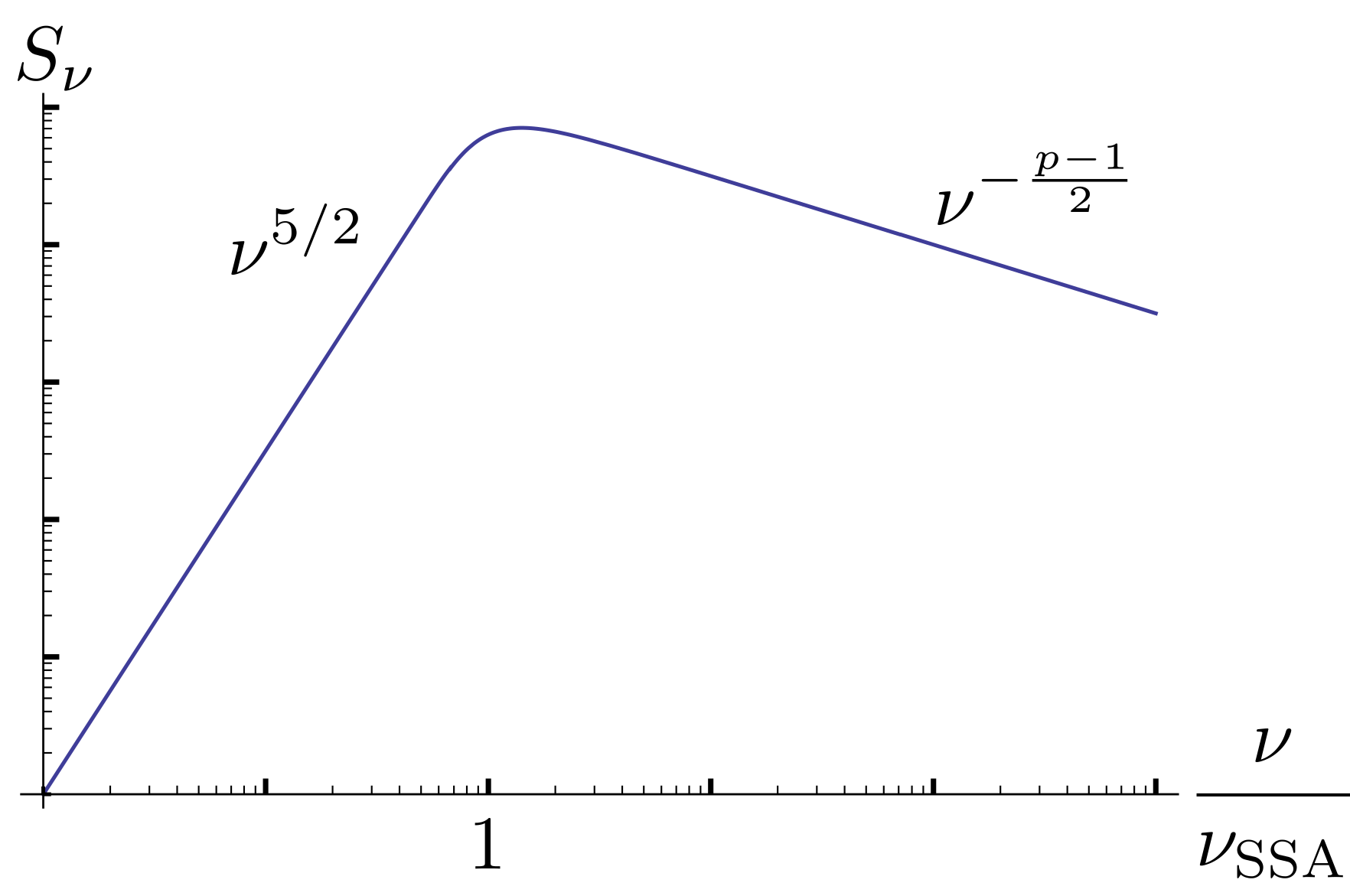}
	\end{center}
	\caption{Sample synchrotron spectrum, with asymptotic behaviors marked.  For optically thick frequencies ($\nu \ll \nu_{\mathrm{SSA}}$) we expect $S_\nu \sim \nu^{5/2}$, and for the optically thin frequencies ($ \nu \gg \nu_\mathrm{SSA}$) we expect $S_\nu \sim \nu^{-\frac{p-1}{2}}$.}
	\label{fig:synchrotronspectrum}
\end{figure}

This spectral shape, can be seen in Figure \ref{fig:synchrotronspectrum}.

\subsubsection{Energetics}
\label{theory:synchrotronenergetics}
In synchrotron emission, energetics provide a convenient way to connect astronomical observables and physical conditions at the source.

The maximal frequency cutoff, $\nu_{max}$ (given in Equation \ref{eq:synchrotronnumaxgammamax}) can set limits on the lifetime of the source.  Not only does Equation \ref{eq:synchrotronfrequencymax} connect the observed frequency cutoff to an inferred electron energy cutoff, but it also predicts that higher energy electrons will radiate energy at a greater rate, depleting their energy more rapidly.  Solving for the time evolution of this system (see \cite{Ginzburg65}) predicts that after a time, $t$, there should be effectively no electrons above a certain energy threshold:

\begin{equation}
	\gamma_{max}(t) < \frac{9}{8 \pi} \frac{1}{\alpha_{FS}} \frac{m_e c^2}{h \nu_g} \frac{1}{\nu_g t}
	\label{eq:electronenergylimit}
\end{equation}
which equivalently places a limit on the time since the most recent electron re-acceleration event:
\begin{equation}
	t < \frac{9}{8 \pi} \frac{1}{\alpha_{FS}} \frac{m_e c^2}{\gamma_{max} h \nu_g} \frac{1}{\nu_g}
	\label{eq:electronagelimit}
\end{equation}

If we assume during that time, $t$, the system has sought a state of minimum total energy (summing kinetic energy of particles in the jet and energy stored in the magnetic field; see Figure \ref{fig:minimumenergy}) we can also determine the minimum-energy magnetic field, $B_{me}$, for a given observation \cite{Worrall2006}:

\begin{figure}[tb]
	\begin{center}
		\includegraphics[width=.7\linewidth]{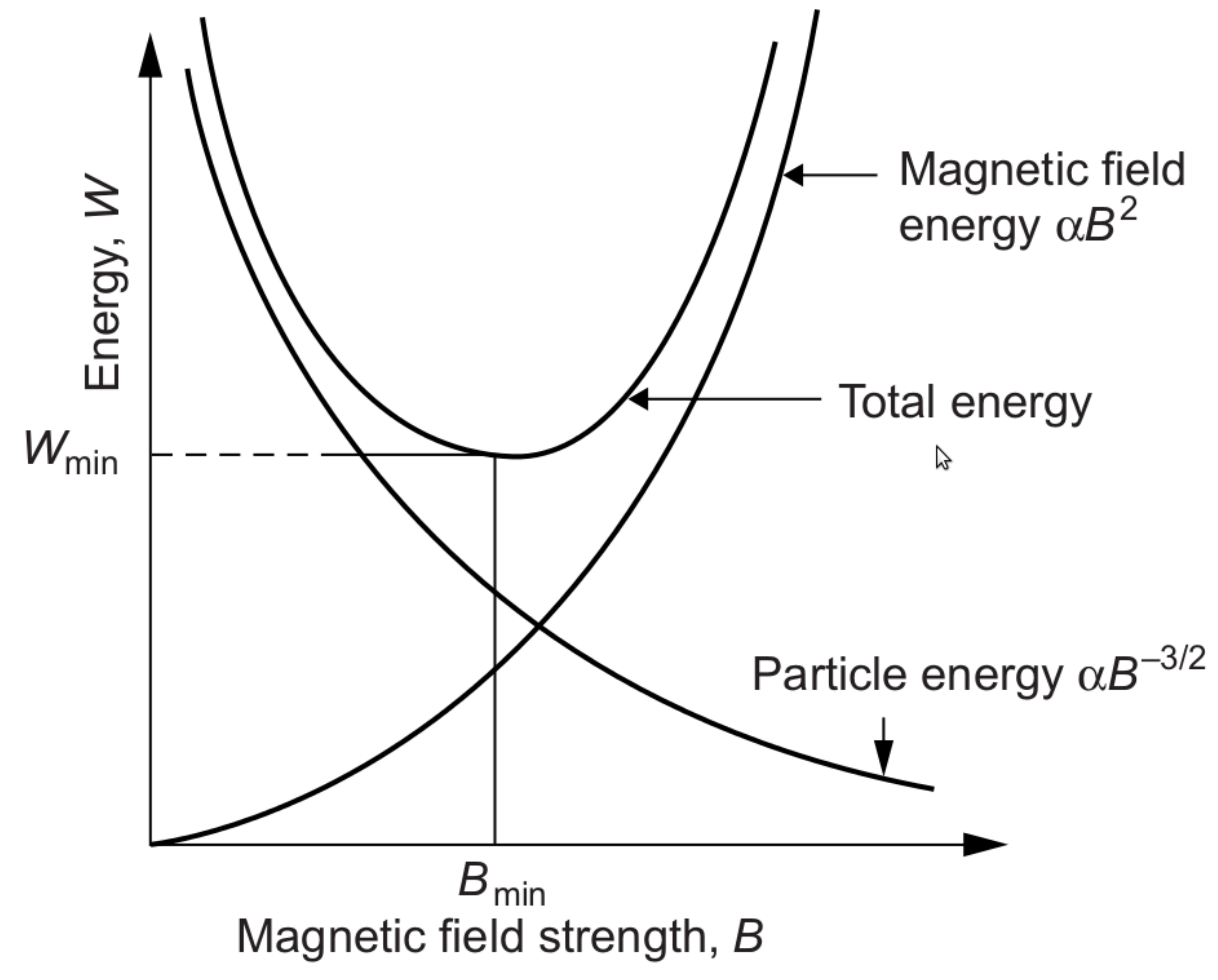}
	\end{center}
	\caption{Sample curve demonstrating the total energy of a synchrotron emitting system, and the magnetic field, $B_{me}$ (marked $B_{min}$ above; see Equation \ref{eq:BmeWorrall} for analytic form), satisfying the minimum energy criterion. Figure taken from \cite{LongairBook3ed}.}
	\label{fig:minimumenergy}
\end{figure}

\begin{equation}
	B_{me} = \left[ \frac{ (p+1) C_1}{4 C_2} \frac{1 + K}{\eta} \frac{4 \pi d_L^2 (1+z)^\frac{p-3}{2}}{V} S_\nu \nu^\frac{p-1}{2} \frac{ \left( \gamma_{max}^{2-p} - \gamma_{min}^{2-p} \right) }{2-p} \right]^\frac{2}{(p + 5)}
	\label{eq:BmeWorrall}
\end{equation}
where $K$ is the ratio of kinetic energy of electrons to the kinetic energy of other particles ($K \approx 0$ gives a ``true'' minimum configuration), $\eta$ is the filling fraction of the emitting volume relative to the relevant volume containing magnetic fields ($\eta =1$ corresponds to no voids, also used for finding a ``true'' minimum), $V$ is simply the volume of the emitting region, and the constants $C_1$, $C_2$ are implicitly defined as:

\begin{eqnarray}
	C_1 &\equiv& \left( \frac{\nu_g}{B} \right)^{-\frac{p-1}{2}} \frac{c}{r_e} \frac{p+1}{3^{p/2} \pi^{3/2}} \frac{\Gamma\left( \frac{p +7}{4}\right)}{\Gamma\left( \frac{p +5}{4}\right)\Gamma\left( \frac{3p +19}{12}\right)\Gamma\left( \frac{3p -1}{12}\right)}
	\label{eq:WorrallC1} \\
	C_2 &\equiv& \frac{u_{mag}}{B^2} = \frac{1}{8 \pi} \:\mathrm{(for \: Gaussian \: cgs)}
	\label{eq:WorrallC2}
\end{eqnarray}

The validity of the assumption of a minimum energy configuration can best be tested through x-rays observations, and an understanding of the information inferred about inverse Compton radiative processes \cite{Worrall09}.  In general, such observations and understandings often, but not always, provide good evidence for a roughly minimal energy configuration \cite{Worrall2006}. To further characterize our system, we must now move on to the theory of \emph{inverse Compton scattering} (\emph{IC}).

\subsection{Inverse Compton Scattering}
\label{theory:InverseComptonScattering}

Another significant source of radiation coming from relativistic plasmas is inverse Compton scattering of photons.  This inelastic scattering is independent of the local magnetic field (in contrast to synchrotron radiation) but requires a source of \emph{seed} photons.  In particular, we'll consider two sources of these photons: external photons (predominantly the cosmic microwave background, \emph{CMB}) and internal photons (such as the source's own synchrotron emission).  We will focus mainly on the analytic theory of external inverse Compton scattering, which will provide a framework that is used numerically to compute synchrotron self-Compton scattering (\emph{SSC}).

Compton scattering is an inelastic scattering process which allows for transfer of momentum and energy between photons and electrons.  In the electron's rest frame, it appears as if an incoming photon has given momentum to the previously stationary electron, increasing the electron's energy.  This process appears qualitatively different when viewed in the center of momentum frame for an ensemble of approximately isotropic electrons and photons.  If the electrons have enough energy, it will appears as if the electrons give energy to the photons.  In \emph{Compton scattering} (in the electron's rest-frame) it appears that the photon loses energy, but in \emph{inverse Compton scattering} (the relativistically equivalent process viewed from the center-of-momentum frame) it appears that the photon has gained energy.

The key technical points are how much energy gets transferred, and often the scattering takes place.  The first is relatively easy to answer.  A low energy photon of frequency $\nu_0$, scattering off an electron of energy $E = \gamma m c^2$, will, on average, result in a photon of frequency: 

\begin{equation}
	\nu \approx \frac{4}{3} \gamma^2 \nu_0
	\label{eq:ComptonBoosting}
\end{equation}

The rate at which this happens depends on the density of photons, the density of electrons, and the cross-section of the interaction.

\subsubsection{Cross-Sections}
\label{theory:InverseComptonCrossSections}
For a single electron, it is most convenient to compute this cross-section in the electron's rest frame.  In that frame, the general result is the \emph{Klein-Nishina cross-section} (derived in \cite{LongairBook2ed}):

\begin{equation}
	\sigma_{\mathrm{K-N}} = \pi r_e^2 \frac{1}{x} \left( \left[1 - \frac{2 \left( x+1 \right)}{x^2} \right] \ln \left( 2x + 1 \right) + \frac{1}{2} + \frac{4}{x} - \frac{1}{2 \left( 2x +1 \right)^2} \right)
	\label{eq:KleinNishinaCrossSectionLongair}
\end{equation}
using the standard definition of the classical electron radius, $r_e \equiv \frac{e^2}{m_e c^2}$, and defining a dimensionless photon energy, $x \equiv \frac{h \nu}{m_e c^2}$.  This can be simplified further in terms of the standard \emph{Thomson cross-section}:

\begin{equation}
	\sigma_T = \frac{8 \pi}{3} r_0^2
	\label{eq:ThomsonCrossSection} 
\end{equation}
which gives us the Equation \ref{eq:KleinNishinaCrossSectionLongair} (in the form given by \cite{RybickiLightmanBook}):
\begin{equation}
	\sigma_{\mathrm{K-N}} =  \frac{3 \sigma_T}{4} \left( \frac{1 + x}{x^3} \left[\frac{2x \left(1 +x\right)}{1 + 2x} - \ln \left(1 + 2x \right) \right] + \frac{\ln \left(1 + 2x\right)}{2x} - \frac{1 + 3x}{\left(1 + 2x\right)^2} \right)
	\label{eq:KleinNishinaCrossSectionRybickiLightman}
\end{equation}
which has the asymptotic behavior:

\begin{equation}
	\sigma_{\mathrm{K-N}} \approx 
	\begin{cases}
		\sigma_T (1 - 2x) \approx \sigma_T & x \ll 1 \\
		\frac{3 \sigma_T}{8} \frac{1}{x} \left[ \ln \left(2x\right) + \frac{1}{2} \right]  & x \gg 1
	\end{cases}
	\label{eq:KleinNishinaCrossSectionLimits}
\end{equation}

(It is important to remember that in the definition of $x \equiv \frac{h \nu}{m_e c^2}$, we care about the photon frequency in the \emph{electron's} rest frame.  While a photon might have a frequency $\nu_0$ in the observer's frame, an electron with energy $E = \gamma m c^2$ could boost that frequency to $\nu = \gamma \nu_0$ in the electron's rest frame.)

Ultimately, we will find that only the Thomson regime plays a significant role in this work (high energy interactions are suppressed due to the $\sigma_\mathrm{K-N} \sim x^{-1}$ dependence at high energies). For that reason we will focus on the details of the low-energy Thomson regime of inverse Compton scattering, where the cross-section is approximately independent of energy.  We leave discussions of the high-energy Klein-Nishina regime to other sources (see \cite{Jones1968,BlumenthalGould1970}).

\subsubsection{General Method for Thomson-Limit Inverse Compton Scattering}
\label{theory:InverseComptonGeneralMethod}
While the technical details of inverse Compton scattering have been extensively covered in other texts (e.g. \cite{BlumenthalGould1970}), we will introduce a simplified approach which we will use to derive results not typically discussed.  The method we introduce has the ability to quickly determine the spectral shape of the scattered radiation, while providing a more intuitive view than the classic texts.

Using the Thomson cross-section, we find the general formula for the production of high energy photons of frequency, $\nu$ (shown through Equations 2.44, 2.45, 2.46 and 2.61 of \cite{BlumenthalGould1970}):

\begin{equation}
	\frac{dn_\nu}{dt} = \int n_\gamma\; \sigma_\mathrm{T} \; c \; n_{\nu_0}\!\!\left(\!\nu_0 = \frac{3 \nu}{4 \gamma^2}\right) d\gamma
	\label{eq:generalICscatteringrate}
\end{equation}
for a distribution of electron energies, $n_\gamma$ (e.g. Equation \ref{eq:ElectronDistributionWithCutoffs}), and an distribution of photon frequencies, $n_{\nu_0}$ (the momenta of each distribution are assumed to be isotropic).

To connect to terms more convenient for radiative transfer, we then find the emission coefficient:

\begin{equation}
	j_\nu = \frac{h \nu}{4 \pi} \frac{dn_\nu}{dt} = \frac{h c \sigma_T}{4 \pi} \nu \int n_\gamma \; n_{\nu_0}\!\!\left(\!\nu_0 = \frac{3 \nu}{4 \gamma^2}\right) d\gamma
	\label{eq:emissioncoefficientgeneral}
\end{equation}

If we assume a homogeneous, optically thin source (see Section \ref{theory:InverseComptonMultipleScatterings} for a relaxation of this assumption), we then find the relation for the shape of the specific flux density:

\begin{equation}
	S_\nu \sim \nu \int n_\gamma \; n_{\nu_0}\left(\nu_0=\frac{3 \nu}{4 \gamma^2}\right) d\gamma
	\label{eq:generalICflux}
\end{equation}
Just as the spectral shape of synchrotron emission was a key observable discussed in Section \ref{theory:synchrotron}, it will be a useful observable when discussing inverse Compton emission.

We can now use Equation \ref{eq:generalICflux}, and assumptions of a homogeneous source, to derive some basic properties of inverse Compton mechanisms within jets.

\subsubsection{Inverse Compton Scattering of the CMB}
\label{theory:InverseComptonCMB}

There are two key regimes when understanding inverse Compton scattering of the cosmic microwave background (\emph{IC-CMB}) in the Thomson limit.  We first will deal with the low frequency tail, which obeys a Rayleigh-Jeans power law, and then we will deal with the higher frequency spectrum, which must take the spectral peak of the CMB black body into account.

\paragraph{Inverse Compton Scattering of a Power Law Photon Distribution}
\label{theory:InverseComptonCMBPowerLaw}

We will first consider the lowest frequencies of the produced spectrum:

\begin{equation}
	\nu \ll \gamma_\mathrm{min}^2 \nu_\mathrm{0, peak}
\end{equation}
where $\nu_{0, \mathrm{peak}}$ is the frequency at the peak of the seed photon distribution.  For a blackbody of temperature, \emph{T}, this peak frequency is given by Wien's law:

\begin{equation}
	\nu_\mathrm{0, peak} = 5.9 \cdot 10^{10} \; T \; \frac{\mathrm{Hz}}{\mathrm{K}}
	\label{eq:WienPeak}
\end{equation}

For those frequencies, we expect every seed photon to come from the Rayleigh-Jeans tail of the blackbody distribution ($\nu_0 \ll \nu_{0, peak}$):
\begin{equation}
	n_{\nu_0} \sim \frac{S_{\nu_0}}{\nu_0} \sim \frac{\nu_0^2}{\nu_0} = \nu_0
	\label{eq:RayleighJeans}
\end{equation}

Using Equations \ref{eq:RayleighJeans} and \ref{eq:generalICflux} we can then find the expected shape of the produced spectrum at low frequencies:

\begin{eqnarray}
	S_\nu &\sim& \nu \int_{\gamma_\mathrm{min}}^{\gamma_\mathrm{max}} n_\gamma(\gamma) \; n_{\nu_0}\!\!\left(\! 3\nu / 4\gamma^2\right)   d\gamma  \nonumber \\
	 &\sim&  \nu  \int_{\gamma_\mathrm{min}}^{\gamma_\mathrm{max}}  \gamma^{-p} \; \frac{\nu}{\gamma^2} d\gamma \nonumber \\
 	 &\sim&  \nu^2  \int_{\gamma_\mathrm{min}}^{\gamma_\mathrm{max}}  \gamma^{-p-2}  d\gamma \\
 	 &\sim&  \nu^2  
	 \label{eq:InverseComptonRayleighJeans}
\end{eqnarray}
(recognizing that $\int_{\gamma_\mathrm{min}}^{\gamma_\mathrm{max}}  \gamma^{-p-2}  d\gamma $ is constant across all frequencies, determined by the index and bounds of the electron energy power law).  At low frequencies, the spectral shape of the Rayleigh-Jeans tail $\left(S_\nu \sim \nu^2 \right)$ is preserved through the inverse Compton scattering.

Above $\nu = \nu_\mathrm{0, peak} \gamma_\mathrm{min}^2$, the Rayleigh Jeans approximation (Equation \ref{eq:RayleighJeans}) no longer holds.  We must account for the peak of the blackbody spectrum.

\paragraph{Inverse Compton Scattering of a Monochromatic Photon Distribution}
\label{theory:InverseComptonCMBMonochromaticseed}

For frequencies, $\nu \gg \nu_\mathrm{0, peak} \gamma_\mathrm{min}^2$, the peak of the blackbody begins to have an effect on the produced spectrum.  Due to the peaked nature of the blackbody spectrum, we now approximate the seed spectrum as monochromatic -- a delta function at $\nu_0 = \nu_\mathrm{0, peak}$.

This derivation is common, and exhaustive discussions can be found in \cite{RybickiLightmanBook,BlumenthalGould1970} (although it is worth noting that the approach using Equation \ref{eq:generalICflux} to find an analog of Equation \ref{eq:InverseComptonRayleighJeans} could also be used to reproduce this well-known result).  Here we will only quote the resulting shape (valid at frequencies $\nu_{0,peak} \gamma_{min}^2 \ll \nu \ll \nu_{0, peak} \gamma_{max}^2$):

\begin{equation}
	S_\nu \sim \nu^{-\frac{p - 1}{2}} 
	\label{eq:InverseComptonMonochromatic}
\end{equation}

Combining the theoretical results of the Rayleigh-Jeans power law component (\ref{eq:InverseComptonRayleighJeans})) and monochromatic component of the IC-CMB spectrum (\ref{eq:InverseComptonMonochromatic}), we then find:

\begin{equation}
	S_\nu \sim
	\begin{cases}
		\nu^2 & \phantom{a \nu_\mathrm{0, peak} \gamma_\mathrm{min}^2 \ll } \nu \ll \nu_\mathrm{0, peak} \gamma_\mathrm{min}^2 \\
		\nu^{-\frac{p - 1}{2}} & \nu_\mathrm{0, peak} \gamma_\mathrm{min}^2 \ll \nu \ll \nu_\mathrm{0, peak} \gamma_\mathrm{max}^2
	\end{cases}
	\label{eq:InverseComptonSpectrum}
\end{equation}

It is important to compare this spectral shape to that of the synchrotron spectrum including self-absorption, Equation \ref{eq:SynchrotronSpectrumFull}.  Both start with a sharp initial rise, the shape of which is independent of the underlying electron distribution.  Then, at higher frequencies, there is a kink, followed by identical photon indices ($S_\nu \sim \nu^{-\alpha}$):

\begin{equation}
	\alpha_\mathrm{IC-CMB} = \frac{p - 1}{2} = \alpha_\mathrm{sync}
\end{equation}

\subsection{Synchrotron Self-Compton}
\label{theory:InverseComptonSynchrotronSelfCompton}
While in Section \ref{theory:InverseComptonCMB} we discussed inverse Compton scattering using the CMB as a seed photon field, many of the same principles can also be applied to inverse Compton scattering of the synchrotron emission produced. As this process requires the same ensemble of electrons to both produce, and then inverse Compton scatter the synchrotron radiation, it is termed \emph{synchrotron self-Compton} radiation (SSC).

The general process is to scatter all local photon fluxes (e.g. starlight, AGN emission, CMB, synchrotron emission, etc); far from any significantly luminous sources, the CMB will often be the dominant source of photons at the relevant radio frequencies.  It is also conceivable that an optically thick source might even inverse Compton scatter photons which have previously undergone inverse Compton scattering.

\subsection{Multiple Scatterings \& Comptonisation}
\label{theory:InverseComptonMultipleScatterings}
Similar to our initial discussion of synchrotron emission, our discussion of inverse Compton scattering has assumed optically-thin inverse Compton scatterings.  For Compton scattering in the Thomson limit (see Section \ref{theory:InverseComptonCrossSections}), we expect a specific optical depth:

\begin{equation}
	\tau_\nu = n_e \sigma_T R
	\label{eq:thomsonopticaldepth}
\end{equation}
where $n_e$ is the number density of electrons:
\begin{equation}
	n_e = \int_{\gamma_{min}}^{\gamma_{max}} n_\gamma d\gamma = \kappa_e \int_{\gamma_{min}}^{\gamma_{max}} \gamma^{-p} d\gamma = \frac{\kappa_e}{1-p} \left( \gamma_{max}^{1-p} - \gamma_{min}^{1-p} \right)
\end{equation}

Ultimately, we will focus little on multiple scatterings, as we find they play little role ($\tau_\nu < 1$). For low energy electrons ($\gamma-1 \ll 1$) the topic of multiple scatterings characterized by the Kompaneets equation (original derived in \cite{Kompaneets1957}; see \cite{LongairBook3ed,RybickiLightmanBook} for suitable summaries). For high energy electrons though (as in our case), we expect the \emph{Compton catastrophe} to drain electron energy extremely rapidly if:

\begin{equation}
	\eta \equiv \frac{L_{IC}}{L_{sync}} = \frac{ \frac{4}{3} \gamma^2 \sigma_T c u_{rad}}{\frac{4}{3} \gamma^2 \sigma_T c u_{mag}} = \frac{u_{rad}}{u_{mag}} > 1
	\label{eq:synchrotroncatastrophe}
\end{equation}
where $u_{rad}$ is the energy density of the photons, and $u_{mag}$ is the energy density of the magnetic field. For more details see \cite{LongairBook3ed}.

\subsection{Bulk Doppler Boosting}
\label{theory:boosting}
In jets, it is not unusual for the previously described synchrotron, SSC and IC-CMB models to be insufficient for modeling observed fluxes.  In particular, the assumption of an isotropic distribution of electrons, whose center-of-momentum frame is the same as the observer's frame, should be suspect since we are dealing with a system undergoing a bulk transfer of mass and power.  If we transform into a frame moving with velocity $v = \beta c$ relative to the observer's frame, the electrons then might be isotropic, undergoing the previously outlined emission processes.  That emission would then be affected by a bulk Lorentz factor $\Gamma = \frac{1}{\sqrt{1 - \beta^2}}$.  When we take into account the angle, $\theta$, between the jet and the line-of-sight (and correspondingly define $\mu \equiv \cos{\theta}$), we get a bulk \emph{Doppler factor}:

\begin{equation}
	\delta = \frac{1}{\Gamma \left(1 - \mu \beta \right)}
	\label{eq:dopplerfactor}
\end{equation}

This bulk Doppler factor is quite noticeable in how it can boost the observed IC-CMB flux relative to the observed synchrotron flux \cite{Worrall09}.  While the Doppler factor, $\delta$, is an observable, the inversion to $\mu$ or $\Gamma$ is degenerate.  We can only choose a fixed $\mu$ and find the required $\Gamma$ (or vice versa) required to explain the observed ratio of synchrotron and IC-CMB fluxes. In particular, it's possible to find the extremal values for $\Gamma$ and $\mu$ which lead to physical values for the other.

Once an assumption for $\Gamma$ or $\mu$ has been made, solving for the other becomes simple using the formalism of \cite{Marshall2005}. Defining a ratio between synchrotron and Compton emission:

\begin{equation}
	R_1 \equiv \frac{S_{\nu_{x-ray}} \nu_{x-ray}^{\alpha}}{S_{\nu_{radio}} \nu_{radio}^{\alpha}}
	\label{eq:MarshallR1}
\end{equation}
where the ``1'' subscript denotes the value that our measurements suggest without the inclusion of Doppler boosting ($\Gamma =1$, hence the notation). We can apply that notation to the magnetic field ($B_{me} \approx B_1 = B \delta$ for $B_{me}$ from Equation \ref{eq:BmeWorrall}), and then group observables into a dimensionless parameter:
\begin{equation}
	K = B_1 \left(a R_1\right)^{\frac{1}{\alpha + 1}} \left( 1 + z \right)^{-\frac{\alpha + 3}{\alpha + 1}}b^{\frac{1 - \alpha }{1 + \alpha}}
	\label{eq:MarshallK}
\end{equation}
with the constants $a$ and $b$, which are given in Gaussian cgs units as: 
\begin{eqnarray}
	a &=& 9.947 \cdot 10^{10} \quad \mathrm{G}^{-2}
	\label{eq:Marshalla} \\
	b &=& 3.808 \cdot 10^{4\phantom{1}} \quad \mathrm{G}
	\label{eq:Marshallb}
\end{eqnarray}

Finally, this allows us to solve for $\Gamma$ or $\theta$ (via $\beta$ or $\mu$ respectively) using Equation 4 of \cite{Marshall2005}:

\begin{equation}
	K = \frac{1 - \beta + \mu - \mu \beta}{\left( 1 - \mu \beta \right)^2}
	\label{eq:MarshallKbetamu}
\end{equation}

\subsection{Extensions of Synchrotron Radiation and Inverse Compton Scattering}
\label{theory:Extensions}

While the theory outlined above is a standard approach for modeling such a system, a more complicated model might be required to describe the emission. We will chiefly discuss additional components to the electron energy distribution, as well deviations from the assumption of isotropic pitch angles, $\alpha_{pitch}$.

We first note that there might be two components to the electron energy power law.  The simplest case is a \emph{kink} in the power law, motivated partially by the electron energy lifetime estimates of Section \ref{theory:synchrotronenergetics} which would preferentially deplete the electron spectrum at high energies.  A more general model would be a two-component electron energy distribution with a distinct low-energy population and a distinct high-energy population (rather than 2 populations which form one continuous energy distribution).  Such a model could explain x-ray emission without appealing to inverse Compton scattering.

It also is possible the electron energy distribution does not follow a power law form.  In particular, \cite{Allen2008} found observational evidence for a ``curved'' electron energy distributions in supernova remnants, which was inspired by theoretical modeling of the shock plasma.  While the theory of the physical processes in jets is not the same as for those in supernova remnants, it does suggest that a more complete handling of the jet physics could yield qualitatively different results, if only we understood the processes within jets well enough.

Finally, the overall assumption of an isotropic distribution of electron momenta was convenient, but could be relaxed.  If electron pitch angles, $\alpha_{pitch}$, aren't correlated with energy, then the distribution of pitch angles only sets the normalization of the synchrotron spectrum, and the location of a synchrotron self-absorption break.  If the pitch angles do correlate with energy (electrons with larger pitch angles tend to radiate faster, so we would expect fewer high energy electrons with large pitch angles) then the modeling becomes significantly more complex, going beyond the scope of this work.

Many of these extensions are best handled numerically, rather than analytically.  Such numerical models are built upon the frameworks which we have developed in this chapter.


\chapter{Methodology}
\label{methodology}

\section{HST Observations}
3 images were proposed for, and obtained, under the HST Guest Observer program (\emph{GO} proposal ID 12261 \cite{HSTProposal}).  These three images were captured with the Wide Field Camera 3 (\emph{WFC3}; for technical details, see the instrument handbook \cite{WFC3InstrumentHandbook}).  In order to gain optical spectral information, three wideband filters were used: F160W, F814W, F475W (centered approximately on 1600nm, 814nm, 475nm respectively; see Appendix A of the instrument handbook \cite{WFC3InstrumentHandbook} for filter details).  The F160W image was taken on the infrared detector of WFC3 (\emph{WFC3/IR}), while the F814W and F475W images were taken on the UV-visible detector (\emph{WFC3/UVIS}).  For more details on the chosen observational parameters, see Table \ref{table:observations}.

\begin{table*}[tbp]\centering
\ra{1.3}
\begin{tabular}{lrrrrrr@{}}\toprule
& & {F160W} & \phantom{abc}& {F814W} & 
\phantom{abc} & {F475W}\\ \midrule 
\emph{Instrument} \\
& WFC3 Detector 							& IR 	& & UVIS & & UVIS \\
&  Native Resolution, [\arcsec]		& .13 	& & .04 & & .04 \\
&  Drizzled Resolution, [\arcsec]		& .07 	& & .02 & & .02 \\

\emph{Filter} \\
& Pivot wavelength, [nm] 						& 1536 	& & 802 & & 477  \\
& Spectral width, [nm] 				& 268 	& & 153 & & 134  \\

\emph{General} \\
& Exposure Time, [s]						& 2708 	& & 1200 & & 1299 \\
& Dataset							& IBJX01010 	& & IBJX01030 & & IBJX01020 \\

\bottomrule
\end{tabular}
\caption{\label{table:observations} Overview of image parameters for relevant HST images. For more technical details on detectors, filters and other instrumental aspects, see \cite{WFC3InstrumentHandbook}. Otherwise all parameters of the observation are similar.}
\end{table*}

\section{HST Pipeline}
\label{methodology:pipeline}
The Space Telescope Science Initiative (\emph{STScI}), on behalf of the National Aeronautics and Space Administration (\emph{NASA}), has developed a standard data calibration pipeline for data taken by the HST.  While for many purposes the default pipeline is sufficient, many users find it more effective to tweak the parameters of this pipelining process.  In this section, we will discuss a few key aspects, as they pertain to this work.

\subsection{Drizzling}
\label{methodology:pipeline:drizzling}
Our work relied heavily on the technique of \emph{drizzling} to harness the full observational power of the HST. Drizzling is a process that takes multiple dithered exposures (images which are slightly offset in their pointing) and combines them to create an image of higher resolution than the native detector image resolution \cite{DrizzleMethod}. Drizzling assumes the image was undersampled -- the resolution must be limited by the detector, not by optical diffraction limits.  As we were hoping to resolve jet substructure, proper drizzling of our images was key to providing the most angular information as possible.

The specific drizzling implementation used in this work was DrizzlePac, which at the time of this writing, was the software supported by the STScI for use with HST \cite{DrizzlePacHandbook}.

\begin{figure}[tb]
	\begin{center}
		\includegraphics[width=.8\linewidth]{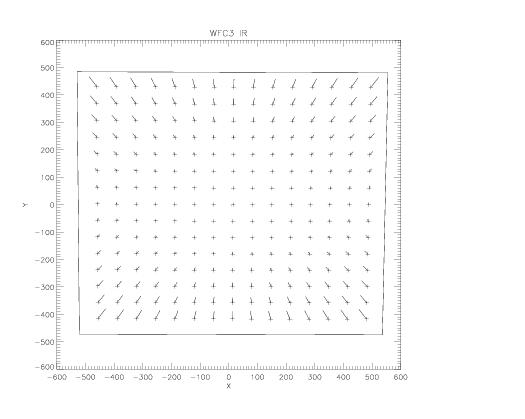}
	\end{center}
	\caption{WFC3/IR geometric distortion map, taken from the WFC3 Data Handbook \cite{WFC3DataHandbook}. Map is in detector pixel coordinates, with crosses representing distortion of angular sizes (scalar distortions) and lines representing vector distortions.}
	\label{fig:geometricdistortions}
\end{figure}

\paragraph{Geometric Distortion Correction}
\label{methodology:pipeline:drizzling:geometricdistortioncorrection}
As drizzling takes a series of exposures, in detector coordinates, and returns a reconstructed image in sky coordinates, it is also a convenient time to correct for geometric distortions inherent to the detector.  Due to instrumental optical effects, the rays reaching the detector are not parallel in all regions \cite{WFC3DataHandbook} (see Figure \ref{fig:geometricdistortions} for a sample distortion map).  This results in a squeezing or stretching of apparent angular sizes.  These distortions are well-characterized, and the drizzling process can invert that distortion, ensuring that all pixels represent equal solid angles of the sky.  The photometry required to determine fluxes involves integrating over solid angles of the sky, thus is it critical that solid angles are properly represented in our images.

\subsection{Cosmic Ray Subtraction}
\label{methodology:pipeline:cosmicraysubtraction}
\begin{figure}[tb]
	\begin{center}
		\includegraphics[width=.6\linewidth]{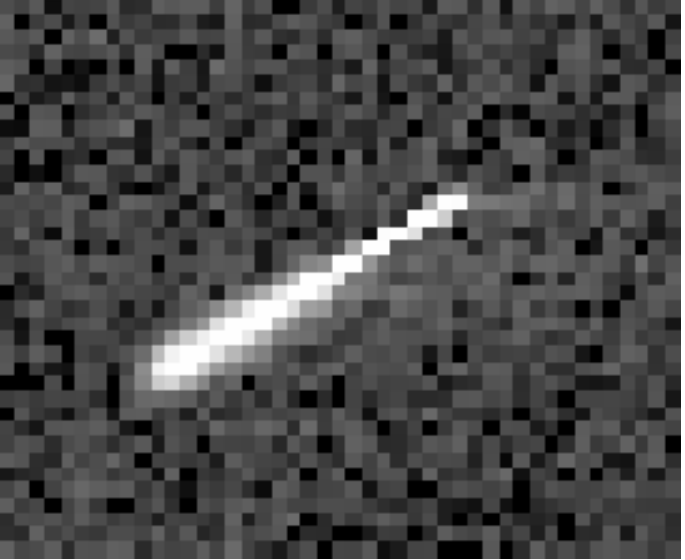}
	\end{center}
	\caption{Sample cosmic ray, taken from a F814W image (pre-drizzle). Notice the high signal-to-noise ratio, with a distinctive narrow track.}
	\label{fig:cosmicray}
\end{figure}

The multiple exposures required for drizzling also allows for the removal of cosmic rays from the images.  Cosmic rays can collide with the space-based detectors, leaving characteristic trails across the image (see Figure \ref{fig:cosmicray}).  Using observed rates of cosmic rays \cite{WFC3InstrumentHandbook}, we could expect as much as 10\% of an image's pixels to be filled with cosmic rays, given the integration times of our images (see Table \ref{table:observations}).  Failure to mask these effects would severely degrade our ability to fit galaxy mophology models and obtain accurate knot fluxes.

Fortunately, cosmic rays are transient, and a single ray should not be seen across multiple exposures.  By searching for significant differences between exposures, and then masking transients which exhibit the distinctive signature of cosmic rays, the image can be cleaned up substantially.  DrizzlePac, in the drizzling process will run a cosmic ray detection and masking algorithm by default, but its parameters should be tweaked as needed (see the DrizzlePac Handbook for more details on the algorithm, and how to ensure that legitimate point sources were not masked \cite{DrizzlePacHandbook}).

\subsection{Error Estimation}
\label{methodology:pipeline:errorestimation}
Finally, after a drizzled image is produced, it is necessary to know the uncertainy associated with the value of each pixel.  While the undrizzled exposures have associated error maps (\verb|err| extensions of \verb|*_flt.fits| files), there is no way to produce an exposure-time weighted drizzle with a correct, corresponding uncertainty map \cite{DrizzlePacHandbook}.

In lieu of rewriting a drizzling algorithm which could produce an exposure-time weighted drizzle while simultaneously producing an accurate uncertianty map, we produced estimated error maps.  This was accomplished using GALFIT (discussed later in Section \ref{methodology:GALFIT}) which combines in quadrature an estimated read noise (constant across the image; estimated using a median test in a background region) and a Poissonian photon noise \cite{GALFITFAQ}.

\section{Galaxy Subtraction through GALFIT}
\label{methodology:GALFIT}

In order to measure the photometric flux of dim knots against a bright background host galaxy, the GALFIT program \cite{GALFIT} was used to fit and remove standard galactic mophology profiles.  At their simplest, these profiles can be Sersic profiles (Eq. \ref{eq:SersicProfile}). For nearby galaxies, it's often necessary to include multiple \Sersic profiles to adequately remove higher order structure within galaxies (for a discussion on adding multiple components, see \cite{GALFITNearby}), and the presence of an AGN can best be modelled through the inclusion of a point source.  To account for instrumental effects, these models were discretized to a grid matching the produced images, and then convolved with the instrument point spread function.

The process of building, convolving and testing a model is then completed for many iterations, searching for best fit model parameters.  The user retains control over what models are being fitted, what parameters are free or fixed, and what constraints are to be enforced on the model parameters.

While GALFIT searches for a local minimum using a $\chi^2$ test, there often exist many local minima.  After a fit is produced, it is critical that a human evaluates the output residual image, changes the initial values of the free parameters, and continues to reiterate GALFIT as needed.

\subsection{HST Point Spread Functions}
\label{methodology:GALFIT:PSFs}

For both modelling a point-like AGN, and for accounting for diffraction effects inherent to the instrumentation, an accurate point spread function (\emph{PSF}) is necessary.  The STScI has created a model PSF creator, TinyTim \cite{TinyTim}, but they note that empirical or extracted PSFs are typically more effective when available \cite{HSTFocus}.  In particular, there are a number of known issues applying TinyTim to the WFC3 instrument \cite{TinyTimErrors}. Furthermore, many of the available methods (see \cite{Starfit,TinyTimErrors}) for fitting model PSF parameters (such as the focus of WFC3) are not meant to handle a PSF at the center of an extended, bright galaxy.

\begin{figure}[htb]
	\begin{center}
		\includegraphics[width=.8\linewidth]{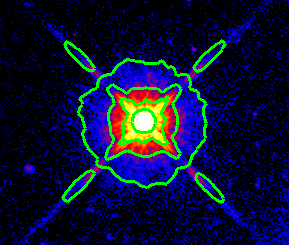}
	\end{center}
	\caption{Extracted PSF, seen through F160W filter, with TinyTim contours overlaid (note particularly the mismatch of the diffraction spike minima).}
	\label{fig:PSF}
\end{figure}

Fortunately, an isolated foreground star was captured in the field-of-view of our images.  That star was cropped and used as an approximate PSF. That star, and a comparison to the TinyTim predicted PSF, can be seen in Figure \ref{fig:PSF}.

\section{Knot Identification}
\label{methodology:knotidentification}
Jet knots were then identified using the processed files resulting from the HST pipelining (Section \ref{methodology:pipeline}) and the galaxy subtraction (Section \ref{methodology:GALFIT}).  Knot candidates were located visually, searching for features which were spatially coincident and similar in morphology to the x-ray and radio jet data (see Figure \ref{fig:XrayRadioMap}).  For each knot candidate, windows bounding the candidate were isolated.  Inside those windows, 2d-gaussians of specific intensity, $I_\nu$, were fit to the observed data, allowing for the knot's semi-major axis to be misaligned with the jet:

\begin{equation}
	I_\nu(x,y) = I_{\nu,0} + I_{\nu,1} \exp\left[ - \frac{ \left(x' - x_0'\right)^2}{2 \sigma_{x'}^2} - \frac{ \left(y' - y_0'\right)^2}{2 \sigma_{y'}^2} \right]
	\label{eq:2dgaussian}
\end{equation}

where:
\begin{eqnarray}
	x' &=& \left(x- x_0\right) \cos\theta - \left(y - y_0 \right) \sin\theta \nonumber \\
	y' &=& \left(x- x_0\right) \sin\theta + \left(y - y_0 \right) \cos\theta \nonumber
\end{eqnarray}

($I_{\nu,0}$, $I_{\nu,1}$, $\sigma_{x'}$, $\sigma_{y'}$, $x_0$, $y_0$, and $\theta$ can all be free parameters in a 2d-gaussian fit.)

The choice of a 2d-gaussian profile is an extension upon a technique previously used on the extragalactic jet PKS-0920 \cite{Marshall01}.  In Section \ref{results:knotphotometries} (specifically Table \ref{table:knotcomparisons}) we show that the 2d-gaussian fit reasonably reproduces the total counts that would be obtained by summing the observed counts (preserving photometric acccuracy), while also yeilding a quantitative measure of the solid angle subtended by the source (determined by $\sigma_{x'}$ and $\sigma_{y'}$). 

Knot candidates were then flagged if their fitted morphologies did not match that of the x-ray or radio data, if the knot did not fully lie on the detector for all images, or if the fit was poor.

For a sample knot image, with best fit gaussian contours overlaid, see Figure \ref{fig:sample_knot_f160w_32arcsec_intro}.

\section{Knot Photometry}
\label{methodology:knotphotometry}
Photometric fluxes from the knots identified in Section \ref{methodology:knotidentification} were then extracted, and the HST calibration was applied to the observed counts, using for the equation:

\begin{equation}
	S_\nu = \frac{DN}{\Delta t} \cdot \mathrm{PHOTFLAM} \cdot \mathrm{PHOTPLAM}^2 \cdot \frac{10^{23}}{2.99 \cdot 10^{18}} \; \mathrm{Jy}
	\label{eq:fluxsensetivity}
\end{equation}
for which $DN$ is the number of observed counts (``data numbers''), $\Delta t$ is the total exposure time, $\mathrm{PHOTFLAM}$ and  $\mathrm{PHOTPLAM}$ are HST-calibrated constants (respectively: the sensetivity of a given instrument through a given filter, and the pivot wavelength of the filter's passband);  the numerical factors are included to give the correct units.

For each window identified in the optical images, the flux over that window was also computed using previously determined x-ray and radio jet profiles (the data reductions were done by Marshall and Lenc for the x-ray and radio data respectively).

\section{Radiative Transfer Modelling}
\label{methodology:radiativetransfer}
Finally, the knot fluxes computed in Section \ref{methodology:knotphotometry} were then used to create numerical spectral models of jet radiative transfer mechanisms.  The analytic theory behind these models can be found in Section \ref{theory:radiativetransfer}; the implementation was handled by the code of Krawczynski \cite{Krawczynski2004}.


\chapter{Results}
\label{results}
The theory and methodology of Chapters \ref{theory} and \ref{methodology} was then applied to images taken from the Hubble Space Telescope (see Table \ref{table:observations} for observing details).  The results of applying those frameworks to the new optical images, when viewed in light of previous x-ray and radio data, are presented in this chapter.

\section{Galaxy Morphological Components}
\label{results:morphologies}
Using the procedures of Sections \ref{methodology:pipeline} and \ref{methodology:GALFIT}, we cleaned, drizzled and galaxy-subtracted the image. As mentioned previously, the process of fitting galaxy models is degenerate, and there are a number of local $\chi^2_\nu$ minima in the parameter space, requiring significant human guidance in setting the initial fit parameters.

Given that freedom, we focused on minimizing residuals at larger radii, rather than focusing on optimally subtracting the AGN core.  This was guided by three insights: the central region would be most susceptible to discrepancies in our PSF, at larger radii any artifacts of the fit would be greatly spread out (appearing as constant offsets at the localized jet knot locations), and the primary x-ray knots (including one which flared) were located at angles greater than 30\arcsec{} from the core.

\begin{figure}[p]
\centering
	\begin{subfigure}{.4\linewidth}
		\includegraphics[width=\textwidth]{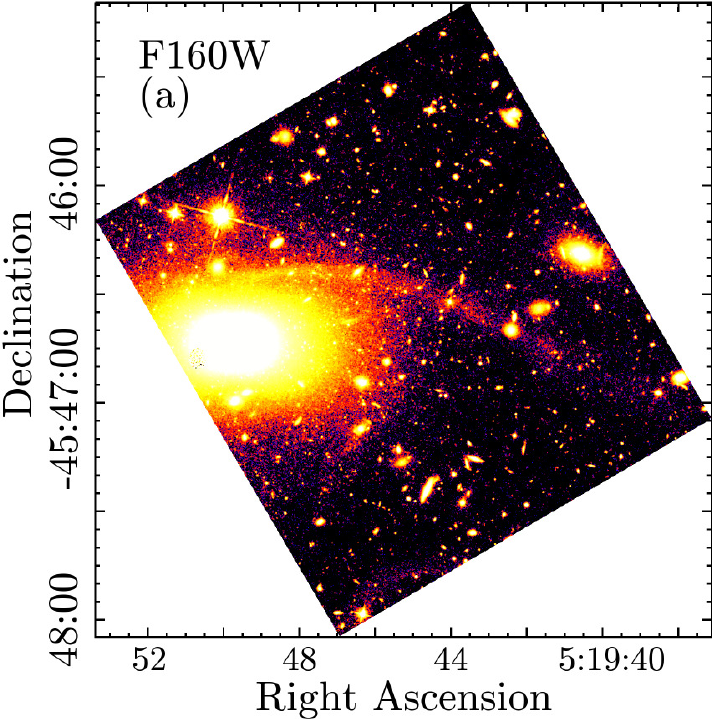}
		\label{fig:galaxysubtractions:f160wgalaxylarge_largefont}
	\end{subfigure}
	\quad \quad
	\begin{subfigure}{.4\linewidth}
		\includegraphics[width=\textwidth]{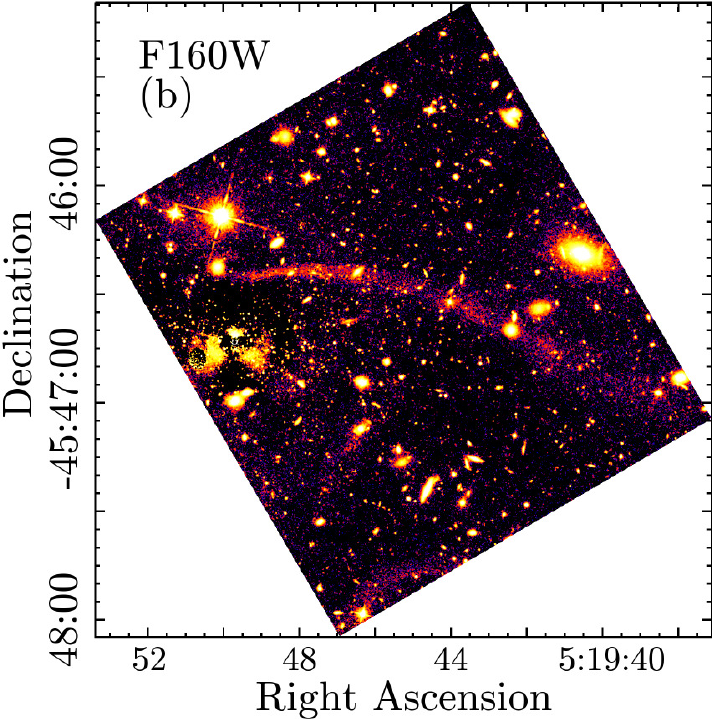}
		\label{fig:galaxysubtractions:f160w_resid}
	\end{subfigure} \\
	\begin{subfigure}{.4\linewidth}
		\includegraphics[width=\textwidth]{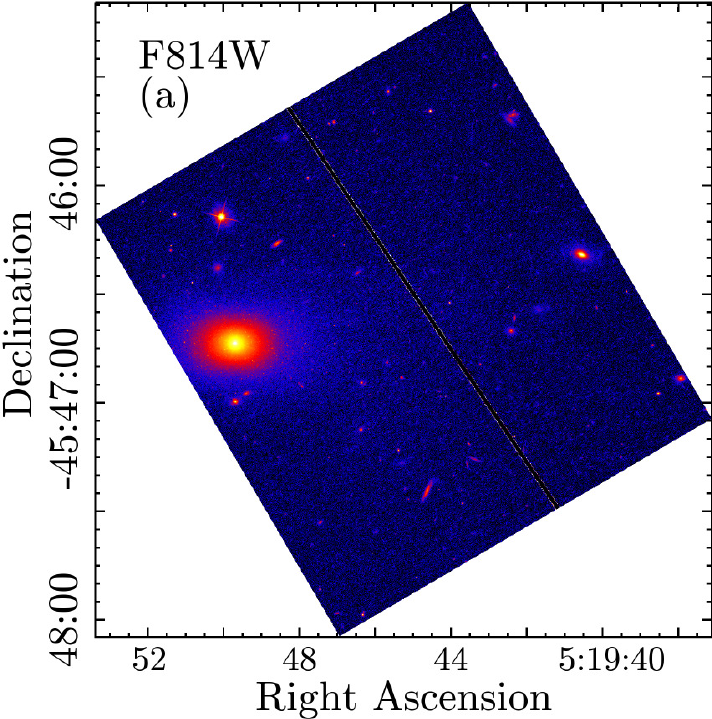}
		\label{fig:galaxysubtractions:f814w_galaxy}
	\end{subfigure}
	\quad \quad
	\begin{subfigure}{.4\linewidth}
		\includegraphics[width=\textwidth]{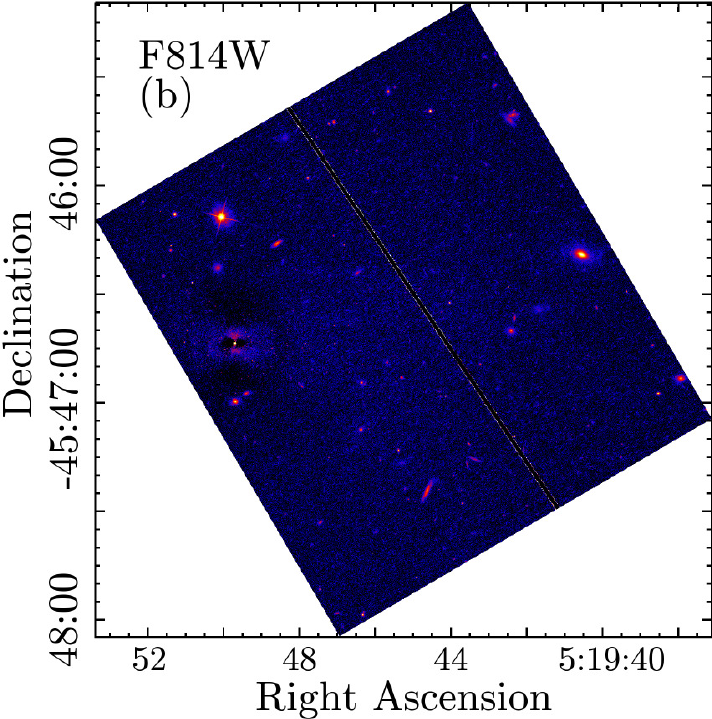}
		\label{fig:galaxysubtractions:f814w_resid}
	\end{subfigure}\\
	\begin{subfigure}{.4\linewidth}
		\includegraphics[width=\textwidth]{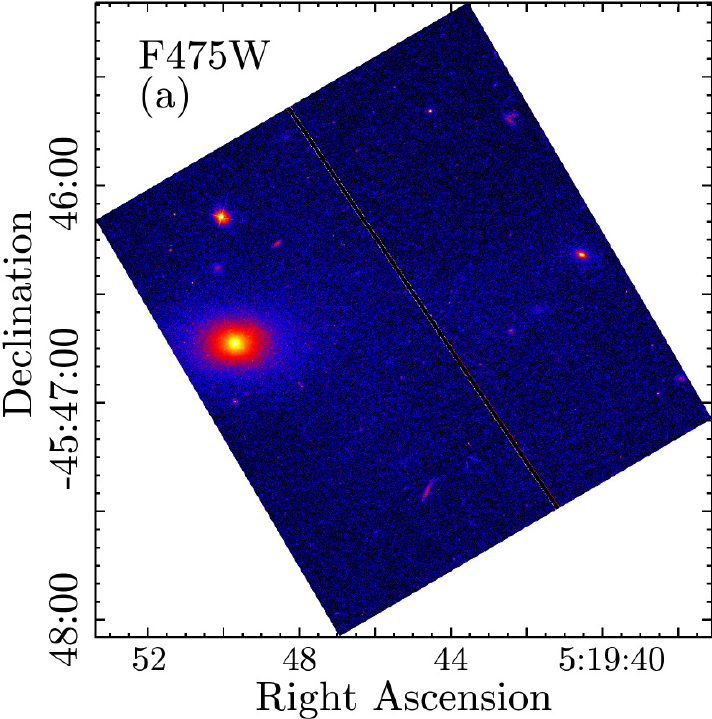}
		\label{fig:galaxysubtractions:f475w_galaxy}
	\end{subfigure}
	\quad \quad
	\begin{subfigure}{.4\linewidth}
		\includegraphics[width=\textwidth]{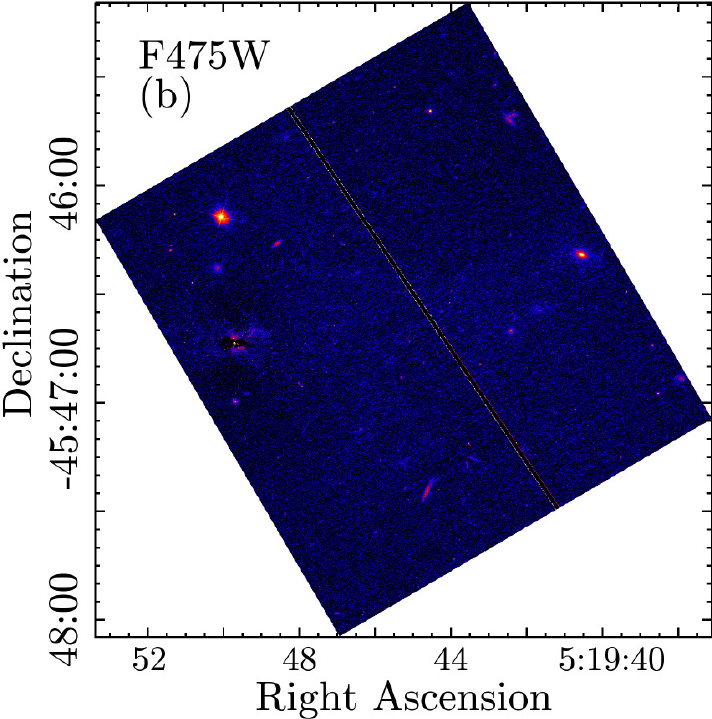}
		\label{fig:galaxysubtractions:f475w_resid}
	\end{subfigure}
	\caption{Images demonstrating the applied galaxy subtractions. (a) images are post-drizzled images, (b) images are the galaxy-subtracted residuals. Note that F160W was taken on a different detector, and for a longer exposure time, leading to the significantly different appearance.}
	\label{fig:galaxysubtractions_largefont}
\end{figure}

To see the galaxy-subtracted images, see Figure \ref{fig:galaxysubtractions_largefont}.

The best-fit parameters found by GALFIT can be seen in Table \ref{table:morphologies}.  For a brief discussion on the family of galaxy models, see Section \ref{theory:morphology}; for a in-depth discussion and motivation for the less standard parameters, see \cite{GALFIT}.

\begin{table*}[p]\centering
\ra{1.3}
\begin{tabular}{lrrrrrr@{}}\toprule
& & {F160W} & \phantom{abc}& {F814W} & 
\phantom{abc} & {F475W}\\ \midrule 

\emph{PSF} \\
& magnitude, $m$ 						& 17.5 	& & 18.0 & & 17.2 \\

\emph{Sky-subtraction} \\
& ADUs/pixel 							& -3.0 	& & .2 & & .2 \\

\emph{Outer bulge} \\
& magnitude. $m$ 						& 16.9 	& & 15.6 	& & 15.5 \\
& \Sersic index, $n$ 					& 2.3 	& & 2.7 	& & 3.1 \\
& effective radius, $R_e$, [\arcsec] 	& 16.6	& & 8.4 	& & 9.2 \\
& axis ratio, $b/a$ 					& .6 	& & .7 		& & .8 \\
& position angle, [\arcdeg]				& -31.7	& & -32.0 	& & -32.5 \\
& diskyness(-)/boxyness(+) 				& 0.5	& &  		& &  \\

\emph{Inner bulge} \\
& magnitude. $m$ 						& 17.4 	\\
& \Sersic index, $n$ 					& 1.6 	\\
& effective radius, $R_e$, [\arcsec] 	 	& 4.2 \\
& axis ratio, $b/a$ 					& .8 	\\
& position angle, [\arcdeg]			& -35.0 \\

\bottomrule
\end{tabular}
\caption{\label{table:morphologies} For more details on each parameter, see \cite{GALFIT}. The F160W image, being a deeper image, required an additional \Sersic bulge to account for the greater range of structure observed.  This does not necessarily reflect a different physical morphology.}
\end{table*}

\section{Knot Photometries}
\label{results:knotphotometries}
\begin{figure}[tbp]
	\begin{center}
		\includegraphics[width=.8\linewidth]{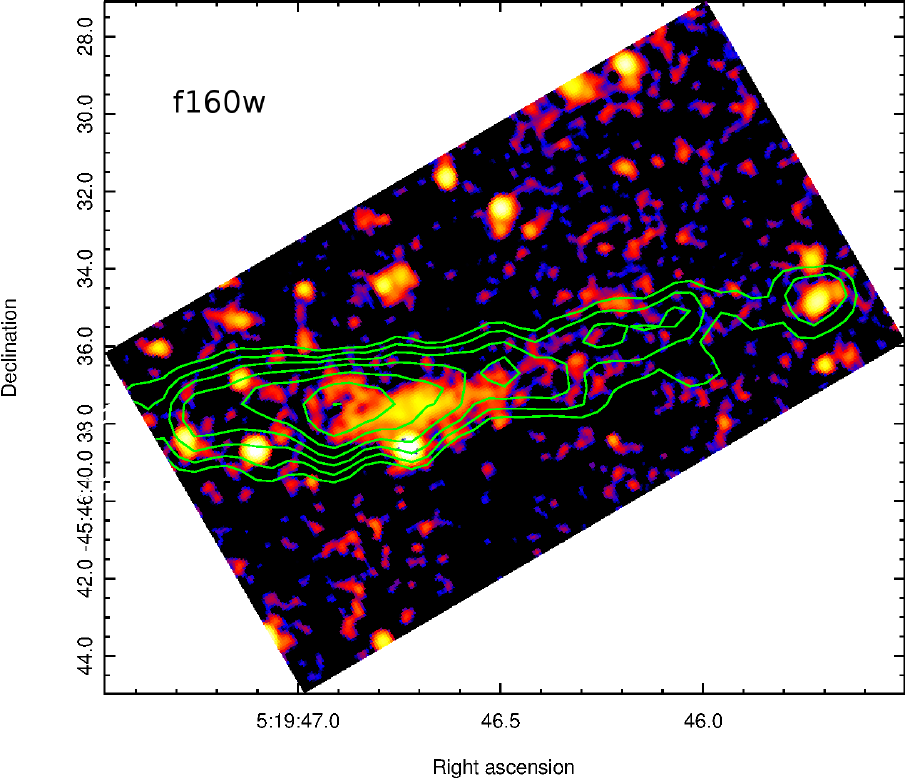}
	\end{center}
	\caption{F160W image of the knot located 32\arcsec{} from the AGN; provided to demonstrate spatial coincidence between x-ray and optical data. Optical data are shown, with x-ray contours overlaid. This is the only knot image with a logarithmic stretch function; all other knot images will have linear stretch functions.}
	\label{fig:knot_xrayoverlay_f160w_32arcsec}
\end{figure}

Using the galaxy-subtracted images (see Figure \ref{fig:galaxysubtractions_largefont}) we looked for features that appeared spatially coincident with the x-ray and radio jets (e.g. Figure \ref{fig:knot_xrayoverlay_f160w_32arcsec}; note that these features are not immediately apparent when viewing the images as a whole).  This led us to discover four knot candidates, at 32\arcsec, 43\arcsec, 106\arcsec{} and 112\arcsec{} from the core. (While those angles allow us to compute a project distance from the core, we have little information about their location relative to the core along the line of sight).

\begin{figure}[tbp]
	\begin{center}
		\includegraphics[width=.8\linewidth]{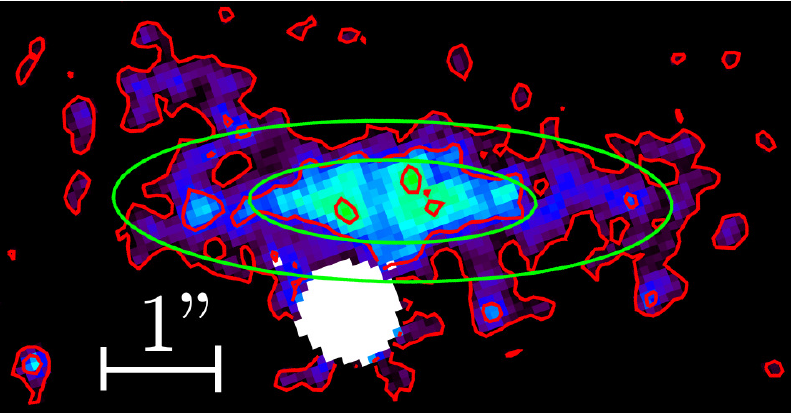}
	\end{center}
	\caption{F160W image of the knot located 32\arcsec{} from the AGN; provided to demonstrate the level of agreement between the data and the 2d-gaussian fits. Observed data are shown both in the image and the red contours; green contours show the best fit. The circular white region is a background galaxy which was masked from the analysis.}
	\label{fig:sample_knot_f160w_32arcsec}
\end{figure}

\begin{figure}[htbp]
	\begin{center}
		\includegraphics[width=.8\linewidth]{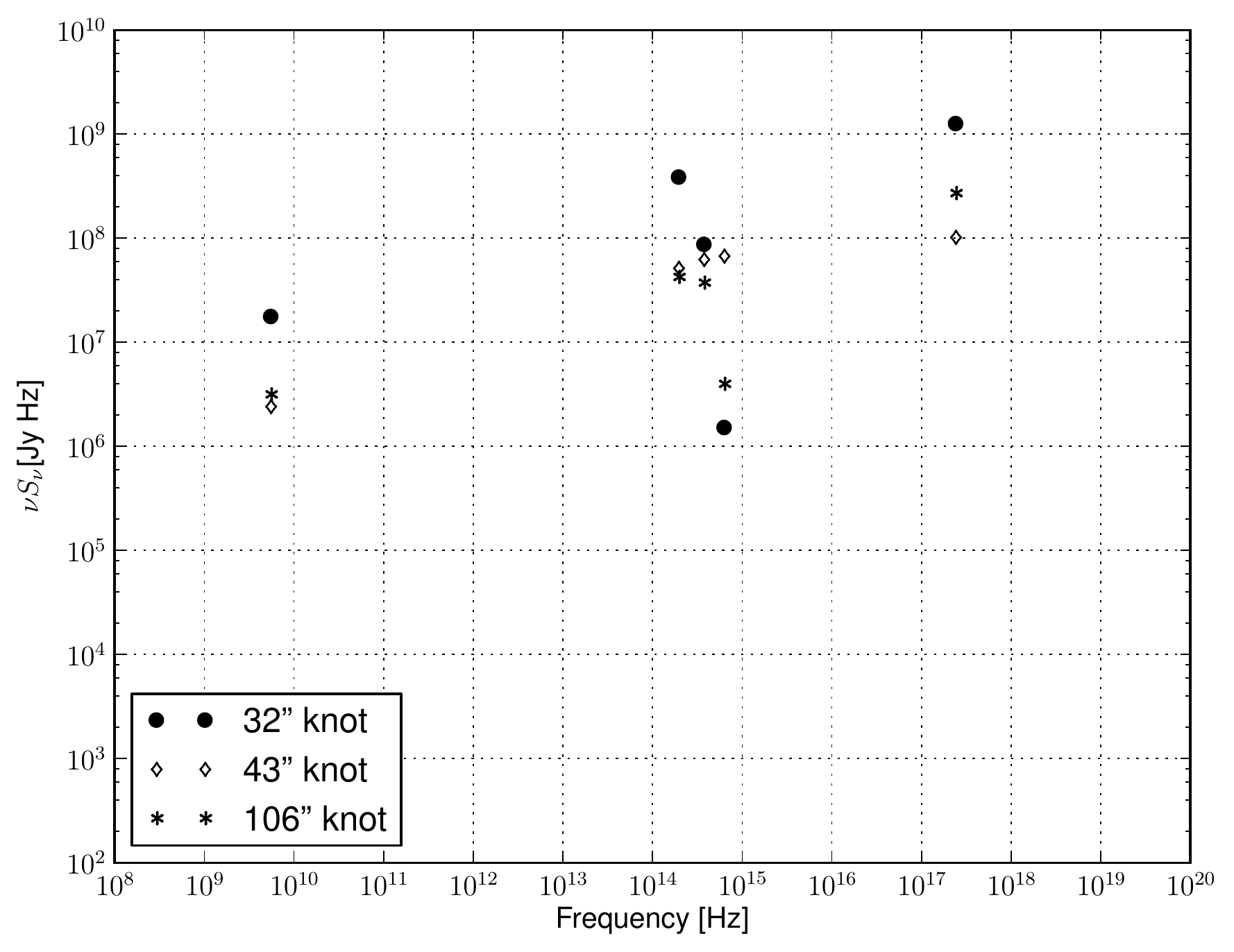}
	\end{center}
	\caption{SED of all knot candidates, except the candidate at 112\arcsec{} (for which we are missing data; 112\arcsec{} knot SED included separately in Figure \ref{fig:SED_nomodels_112arcsec}). Error bars have been omitted, as the error is typically smaller than the markers or the points could only be treated as upper limits for any detectable flux.}
	\label{fig:SED_nomodels_allknots}
\end{figure}

By fitting 2d-gaussians to these knots (see Equation \ref{eq:2dgaussian}) we extracted quantitative measures of knot sizes and fluxes (see Figure \ref{fig:sample_knot_f160w_32arcsec} for an example, Table \ref{table:fittedknots} for reduced photometry results and Appendix \ref{appendix:knotcandidateimages} for all knot candidate images).  These knots consistently had the most flux in the F160W images (as opposed to the F814W and F475W images), although in many cases we could only determine upper limits for the F814W and F475W knot candidate fluxes.  These fluxes were then checked by simply summing the detected counts across the entire image.  This secondary method avoids the systematic biases introduced by assuming a 2d-gaussian profile, and allows for a more direct comparison to the radio and x-ray data (produced by Lenc and Marshall, respectively, who collapsed the transverse data about the jet, leaving only longitudinal information, making a 2d-gaussian fit impossible).  As both techniques, 2d-gaussian fitting and simple summing, produce similar results (see Table \ref{table:knotcomparisons}), we feel the data are sufficiently robust.  A spectral energy distribution (\emph{SED}) of the observed fluxes can be found in Figure \ref{fig:SED_nomodels_allknots}. (Note: observed fluxes are conventionally reported in Janskys, Jy, such that 1 Jy $ = 10^-23$ erg s$^{-1}$ cm$^{-2}$  Hz$^{-1}$.)

\begin{figure}[tb]
	\begin{center}
		\includegraphics[width=.8\linewidth]{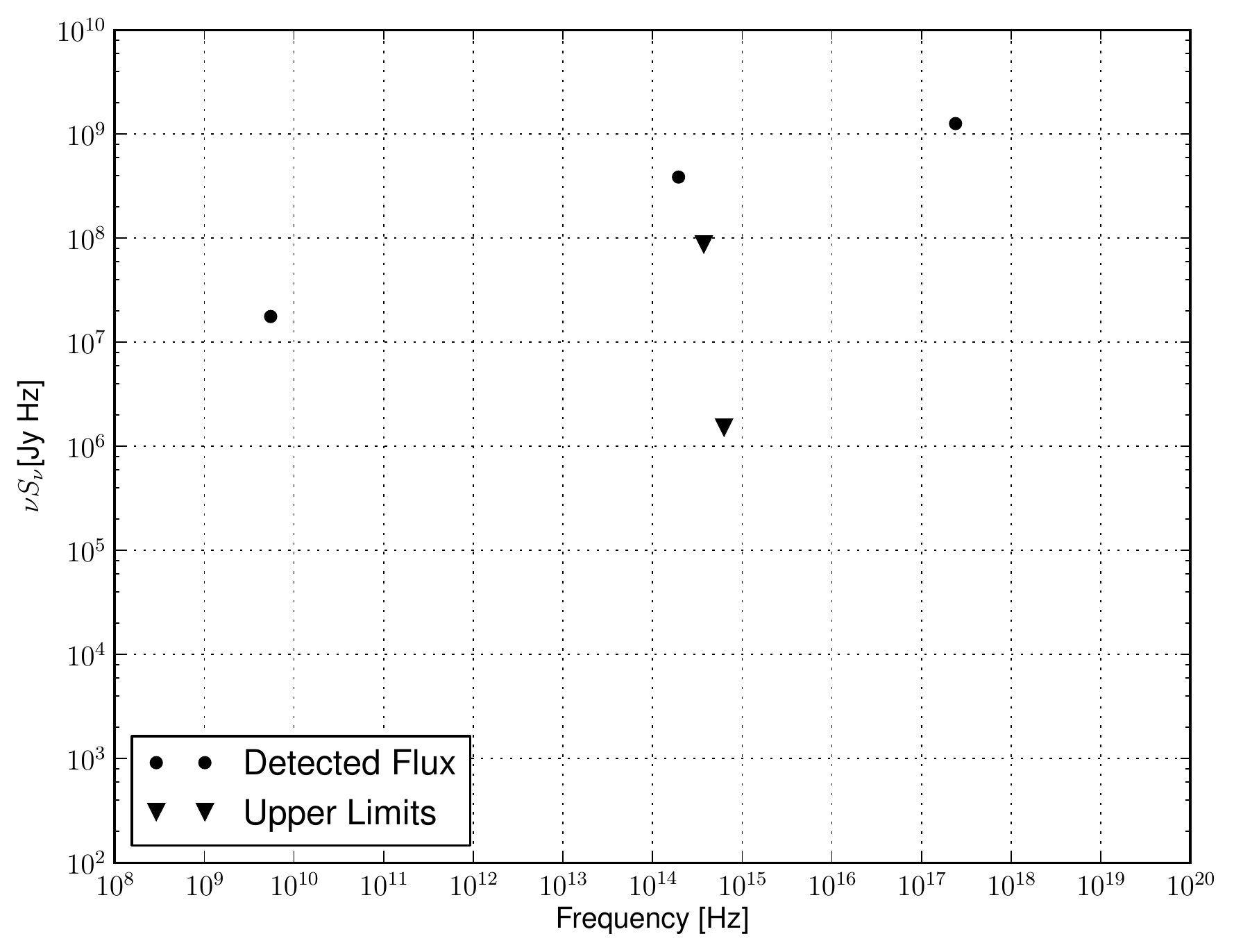}
	\end{center}
	\caption{SED of data only from the 32\arcsec{} knot. Flux uncertainties are smaller than the markers, and thus have been omitted.}
	\label{fig:SED_nomodels_only32arcsec}
\end{figure}

One final oddity is the discrepancy between F160W and F814W images of the knot observed at 32\arcsec{} away from the core.  While the F160W source is clearly extended, we observe a fully unresolved point source in the F814W images (it has a full width at half maximum, \emph{FWHM}, of .1\arcsec, approximately equal to the expected PSF FWHM \cite{WFC3InstrumentHandbook}).  In many ways this was unexpected: the canonical view of a jet holds that the luminosity profile should not change significantly with small changes in frequency.  Furthermore, as a point source, it has a greater intensity but a lower flux than the extended F160W source -- it takes a lower flux and concentrates it within a smaller solid angle on the sky. These results are unusual, and we flag the F814W knot for further analysis.  It is worth noting that very little of our analysis would be changed even if chose to remove this point source from our analysis -- Figure \ref{fig:SED_nomodels_only32arcsec} clearly shows a drop in flux between F160W and F475W, whether or not we include the F814W point source.

For the rest of this analysis we will focus on the 32\arcsec{} knot.  This knot corresponds to the brightest x-ray knot \cite{Wilson01}, and was detected with the highest statistical significance (the null hypothesis was ruled out with greater than $5\sigma$ confidence), and does not lie on the edge of any detector chips (unlike the 112\arcsec{} knot in the F160W image).  An SED of just 32\arcsec{} knot can be seen in Figure \ref{fig:SED_nomodels_only32arcsec}.

\begin{table*}[p]\centering
\ra{1.3}
\begin{tabular}{lrrrrrr}\toprule
& & {F160W} & \phantom{abc}& {F814W} & 
\phantom{abc} & {F475W}\\ \midrule

\emph{32\arcsec} \\
& counts					& 40000 & & $<$650 	& & $<$40 	\\
& flux [$\mu$Jy] 			&  2.2	& & $<$.18	& & $<$.01	\\
& FWHM (x) [\arcsec] 		& 3.3 	& & .02		& &  .13	\\
& FWHM (y) [\arcsec] 		& .9 	& & .02		& &  .02	\\
& rotation [\arcdeg] 		& 1.1 	& & 0.0		& &  25.5	\\

\emph{43\arcsec} \\
& counts					& 6800 	& &  $<$650	& & $<$740 	\\
& flux [$\mu$Jy] 			&  .38	& & $<$.18	& & $<$.11 	\\
& FWHM (x) [\arcsec] 		&  .30	& & .28 	& & .24	\\
& FWHM (y) [\arcsec] 		&  .45	& & .17 	& & .18	\\
& rotation [\arcdeg] 		&  51.9	& & -25.0 	& & -40.6\\

\emph{106\arcsec} \\
& counts					& 3700 	& & $<$360 	& & $<$80 	\\
& flux [$\mu$Jy] 			&  .21	& &  $<$.10	& & $<$.01 	\\
& FWHM (x) [\arcsec] 		&  .31 	& & .19		& & .13 	\\
& FWHM (y) [\arcsec] 		&  .35 	& & .11 	& & .09 	\\
& rotation [\arcdeg] 		& 37.7 	& & -14.9 	& & 0.0 	\\

\emph{112\arcsec} \\
& counts					& 68000 & & 8200 	& & 8300 	\\
& flux [$\mu$Jy] 			&  3.8	& & 2.2 	& & 1.2	\\
& FWHM (x) [\arcsec] 		& 1.2 	& & .85 	& & 1.2 	\\
& FWHM (y) [\arcsec] 		& .57 	& & .43 	& & .42 	\\
& rotation [\arcdeg] 		& -.2 	& & 1.2 	& & 6.0	\\

\bottomrule
\end{tabular}
\caption{\label{table:fittedknots} Results of 2d-gaussian fits for observed optical knot candidates.  The (x) and (y) components of the FWHM correspond to parallel to the jet and perpendicular to the jet, respectively, with a misalignment given by \emph{rotation} (which comes from fitting 2d-gaussians with elliptical contours; (x) and (y) are the semi-major and semi-minor axes). While the 112\arcsec{} knot might have been the strongest detection, it lies on the edge of the F160W detector and was removed from further analyses.}
\end{table*}

\begin{table*}[tbp]\centering
\ra{1.3}
\begin{tabular}{llrrrrr}\toprule
& & {F160W} & \phantom{abc}& {F814W} & 
\phantom{abc} & {F475W}\\ \midrule

\emph{32\arcsec  } & \\
flux [$\mu$Jy]&  	 fitted &  2.24	& & $<$.18	& & $<$.01	\\
& 				 	 summed &  1.98	& & $<$.23	& & $<$.01	\\

\emph{43\arcsec  } & \\
flux [$\mu$Jy]&  	 fitted &  .38	& & $<$.18	& & $<$.11	\\
& 				 	 summed &  .26	& & $<$.17	& & $<$.11	\\

\emph{106\arcsec  } & \\
flux [$\mu$Jy]&  	 fitted &  .21	& & $<$.10	& & $<$.01	\\
& 				 	 summed &  .22	& & $<$.10	& & $<$.01	\\

\emph{112\arcsec  } & \\
flux [$\mu$Jy]&  	 fitted &  3.80	& & 2.24	& & 1.25	\\
& 				 	 summed &  		& & 2.47	& & 1.14	\\ 

\bottomrule
\end{tabular}
\caption{\label{table:knotcomparisons} Comparison of knot flux extraction methods.  ``Fitted'' fluxes refer to fitting a 2d-gaussian to the data, and ``summed'' fluxes refer to simply summing the counts within a given window. For more details, see Section \ref{methodology:knotphotometry}. Summed flux is missing from the F160W knot at 112\arcsec, since that knot extends past the edge of the image.}
\end{table*}

\section{Spectral Energy Distribution Models}
\label{results:seds}

\subsection{Analytic Results}
\label{results:seds:AnalyticResults}
Using the observed knot fluxes seen in Figure \ref{fig:SED_nomodels_only32arcsec} (listed in Table \ref{table:knotcomparisons}), along with data from previous studies \cite{Hardcastle05, Marshall10, Perley97,Tingay08,Wilson01}  we can begin to place quantitative constraints on jet emission mechanisms.  First, we note that a unbroken power law SED cannot explain all of the observations (see Figure \ref{fig:SED_nomodels_only32arcsec}).  The next most commonly applied model is a 2-mechanism model, produced by a single population of electrons (e.g. synchrotron emission at low frequencies, inverse Compton scattering at high frequencies). We will focus mainly on inverse Compton scattering of the cosmic microwave background (\emph{IC-CMB}), simply noting that inverse Compton scattering of synchrotron photons (\emph{SSC}) is numerically shown to be less significant (see Figure \ref{fig:SEDS_p100_p150_p200_p288}).

Assuming a single power law population of electrons (Equation \ref{eq:ElectronDistributionWithCutoffs}) allows us to connect synchrotron observables to inverse Compton observables, placing constraints on both.  In particular, theory predicts that the spectral index of synchrotron radiation, $\alpha_{sync}$, be identical to the spectral index for high energy IC-CMB emission, $\alpha_{IC-CMB}$, when far away from the relevant $\gamma_{min}$, $\gamma_{max}$ cutoffs.

While Wilson \cite{Wilson01} and Hardcastle \cite{Hardcastle05} found jet x-ray photon indices of $\Gamma_{x-ray} = 1.94^{+ .43}_{-.49}$ and $\Gamma_{x-ray} = 1.97 \pm .07$ (where $\alpha = \Gamma_{x-ray} - 1$ \cite{XrayGamma} and $S_\nu \sim \nu^{- \alpha}$), those values were determined by combining data for the entire jet.  Such an approach would not be appropriate when considering the optical knots -- most of the jet is undetectable at optical frequencies, so we are limited to studying the abnormally luminous regions.  We will leave the x-ray spectral index, $\alpha$ (and thus the electron energy index, $p$) unconstrained, and will note the significance of deviations from the results of \cite{Wilson01,Hardcastle05} in Chapter \ref{significance}.

\begin{figure}[tb]
	\begin{center}
		\includegraphics[width=.8\linewidth]{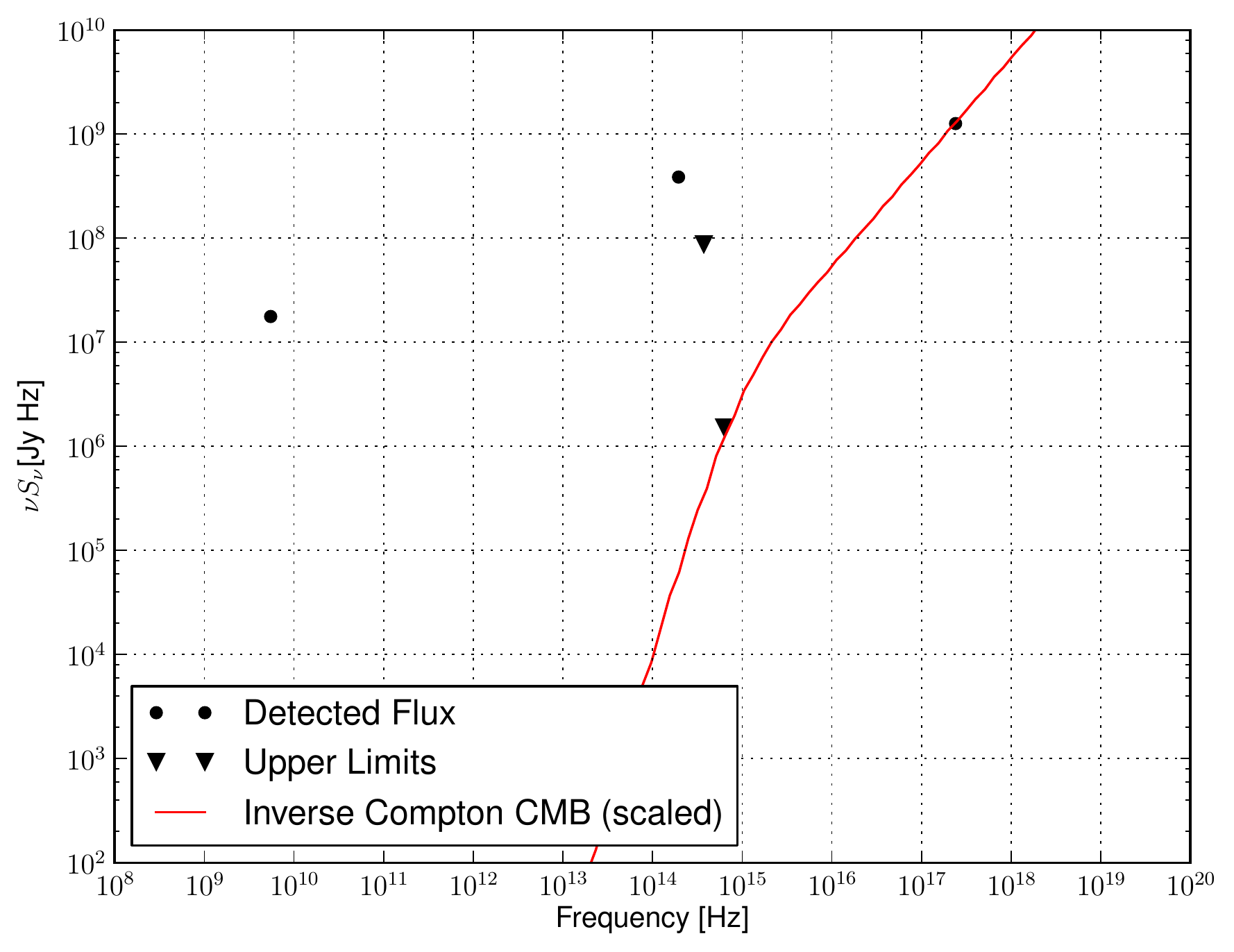}
	\end{center}
	\caption{SED demonstrating the minimal $\gamma_{min}$ constraint, using the IC-CMB cutoff. Modeled IC-CMB emission (red) has been normalized to match the observed x-ray flux, since we are only currently concerned with the spectral shape and location of the cutoff.}
	\label{fig:SED_gamma_min_sample}
\end{figure}

For a given electron index, $p$, we can solve for the predicted values of a number of physical parameters. First, we can constrain $\gamma_{min}$ using location of the low-energy cutoff of IC-CMB emission. Assuming that the IC-CMB SED is above it's low energy cutoff by the time it reaches the frequencies of the observed x-ray emission, we can solve for the smallest $\gamma_{min}$ that can both explain an optical non-detection, and x-ray emission matching our observations (see Table \ref{table:estimatedvalues} for results, and Figure \ref{fig:SED_gamma_min_sample} for an example).  Next, we can simultaneously solve for $\gamma_{max}$ and $B_{me}$ using Equations \ref{eq:synchrotronnumaxgammamax} and \ref{eq:BmeWorrall}, using the F160W frequency as $\nu_{max}$ (for the rest of the analysis, we assume $B = B_{me}$ as motivated by \cite{Worrall09}).  Knowing $B$ and $\gamma_{max}$ lets us solve for an upper limit on the age of the electron distribution, $t_{age}$ (Equation \ref{eq:electronagelimit}), which we find to be comparable to, but typically greater than the light-crossing time assuming an projected geometry without Doppler boosting. (Electron age limits are a weak constraint, as they do not account for continuing processes; they simply demonstrate physical consistency in the understanding which we are building.)  Next, knowing the electron energy cutoffs allows us to find the normalization of the electron energy distribution, $\kappa_e$ which can be determined through Equations \ref{eq:gouldfluxspectrum} and \ref{eq:gouldrhonu}.  Finally, those parameters predicts a synchrotron self-absorption frequency, $\nu_{SSA}$, given by Equation \ref{eq:nuSSA}, which must agree to the limits required for a SSA model (with a free $\nu_{SSA}$) to match the radio and optical data.

The results of these calculations, computed for a range of electron indices, $p$, can be found in Table \ref{table:estimatedvalues}. 

\begin{table}[p]\centering
\ra{1.3}
\begin{tabular}{lllrrrrlrrlrrrrrr}\toprule
$p$  & \phantom{a}& $\alpha = \frac{p-1}{2}$ & $\gamma_{min}$ &  $\gamma_{max}$  & $B_{me}$ & \phantom{a}& $\kappa_e$ &  $\nu_{SSA}$   &  \phantom{a}& & $t_{age}$\\ 
  && & & $10^6$ & $\mu$G && $\mathrm{cm}^{-3}$ &  Hz  && &$10^3$ yr\\ 
  && & &        &        &&                    & \emph{OD}  && \emph{SED} \\ 

\midrule

1.0  	&& 0.00 	& 60  & 2 & 39  && $2 \cdot 10^{-12}$ & $3 \cdot 10^{4}$ && $<5 \cdot 10^{9} $ & 8 \\ 
1.5  	&& 0.25 	& 80  & 2 & 26  && $4 \cdot 10^{-10}$ & $1 \cdot 10^{5}$ && $<5 \cdot 10^{9} $ & 14 \\ 
2.001	&& 0.5005	& 110 & 3 & 21  && $6 \cdot 10^{-8}$  & $3 \cdot 10^{5}$ && $<5 \cdot 10^{9} $ & 20 \\ 
2.5  	&& 0.75 	& 130 & 2 & 26  && $5 \cdot 10^{-6}$  & $7 \cdot 10^{5}$ && \phantom{$<$ }$6 \cdot 10^{9}$ & 14 \\ 
3.0  	&& 1.00 	& 160 & 2 & 37  && $2 \cdot 10^{-4}$  & $1 \cdot 10^{6}$ && \phantom{$<$ }$1  \cdot 10^{10}$ & 8 \\ 
3.5  	&& 1.25 	& 180 & 2 & 51  && $8 \cdot 10^{-3}$  & $3 \cdot 10^{6}$ && \phantom{$<$ }$3  \cdot 10^{10}$ & 5 \\ 
4.0  	&& 1.50 	& 200 & 2 & 68  && $3 \cdot 10^{-1}$  & $5 \cdot 10^{6}$ && \phantom{$<$ }$4 \cdot 10^{10}$ & 3 \\ 
4.5  	&& 1.75 	& 220 & 1 & 85  && $1 \cdot 10^{1}$   & $8 \cdot 10^{6}$ && \phantom{$<$ }$7 \cdot 10^{10}$ & 2 \\ 
5.0  	&& 2.00 	& 250 & 1 & 103 && $3 \cdot 10^{2}$   & $1 \cdot 10^{7}$ && \phantom{$<$ }$1 \cdot 10^{11}$ & 2 \\

 \\
1.90  	&& 0.45 	& 100 & 3 & 21 && $2 \cdot 10^{-8}$ & $2 \cdot 10^{5}$  && \phantom{$<$ }$2 \cdot 10^{9}$ & 19 \\
2.88  	&& 0.94 	& 150 & 2 & 34 && $9 \cdot 10^{-5}$ & $1 \cdot 10^{6}$  && \phantom{$<$ }$1 \cdot 10^{10}$ & 10 \\ 
3.74  	&& 1.37 	& 190 & 2 & 57 && $5 \cdot 10^{-2}$ & $4 \cdot 10^{6}$  && \phantom{$<$ }$3 \cdot 10^{10}$ & 4 \\

\bottomrule
\end{tabular}
\caption{\label{table:estimatedvalues} Estimates of key physical values, as outline in Section \ref{theory:radiativetransfer}. A representative sample of electron indices, $p$, have been chosen, followed by the electron indices suggested by the data of Wilson \cite{Wilson01} ($p=2.88_{-.90}^{+.86}$, assuming $p = 2 \alpha + 1$).  Also note that $p=2$ is a special case not handled by many of the equations used (e.g. Equation \ref{eq:BmeWorrall}), so $p \approx 2$ has been used instead as a representative value. \\ For $\nu_{SSA}$, ``OD'' refers to the estimate predicted by analytic expressions of optical depth (Equation \ref{eq:synchrotronopticaldepth}); ``SED'' denotes the frequency that would be required for a model to match the observed radio and F160W fluxes (the two columns of $\nu_{SSA}$ must match for a self-consistent model). Finally, remember that $t_{age}$ sets only an upper limit on the time since electron re-acceleration, not the actual age of the electron population (see Equation \ref{eq:electronagelimit}).}
\end{table}

\subsection{Numerical Results}
\label{results:seds:NumericalResults}
\begin{figure}[p]
\centering
	\begin{subfigure}{.49\linewidth}
		\includegraphics[width=\textwidth]{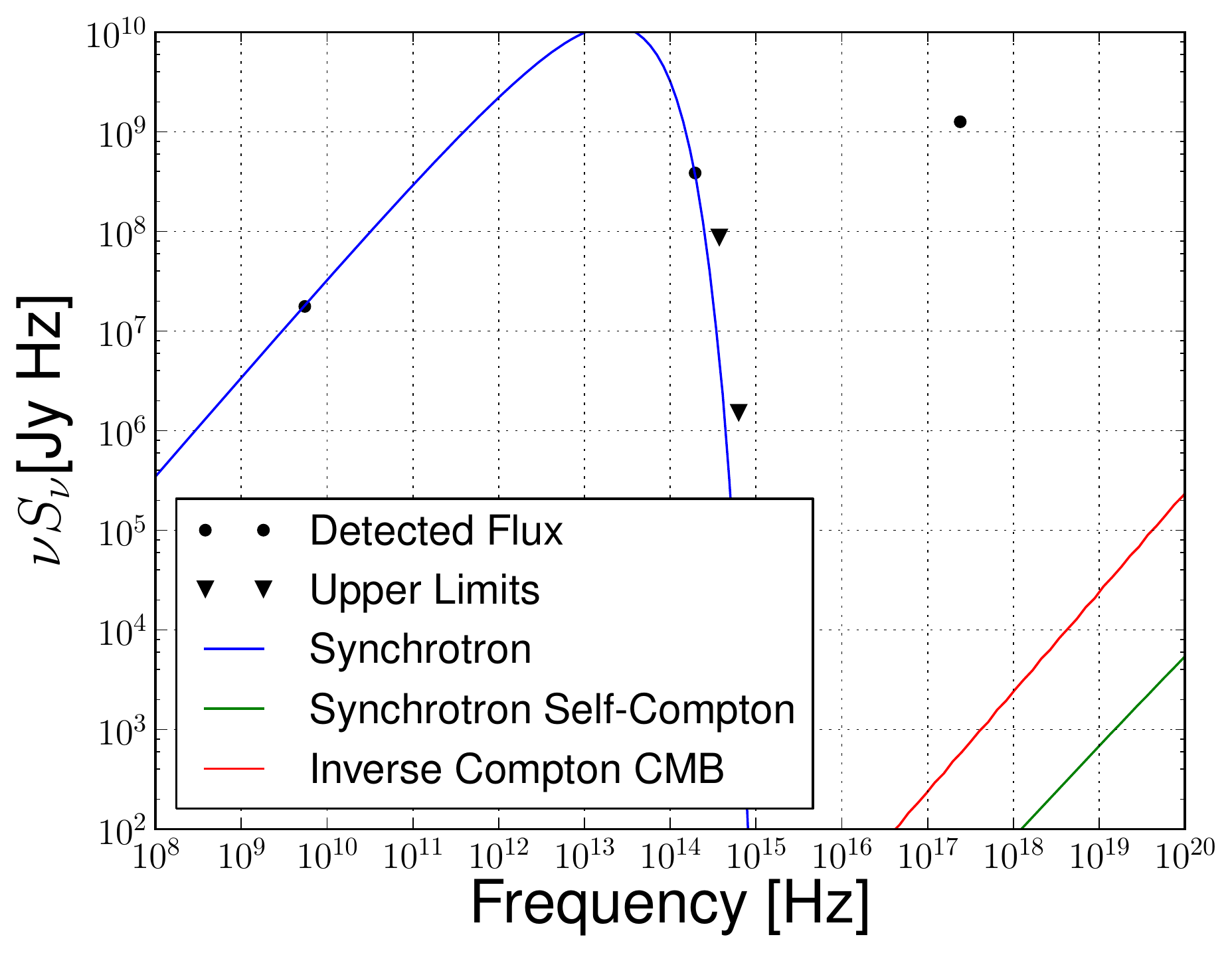}
		\caption{\label{fig:galaxysubtractions:SEDS_p100} $p=1.00$ }
	\end{subfigure}
	\begin{subfigure}{.49\linewidth}
		\includegraphics[width=\textwidth]{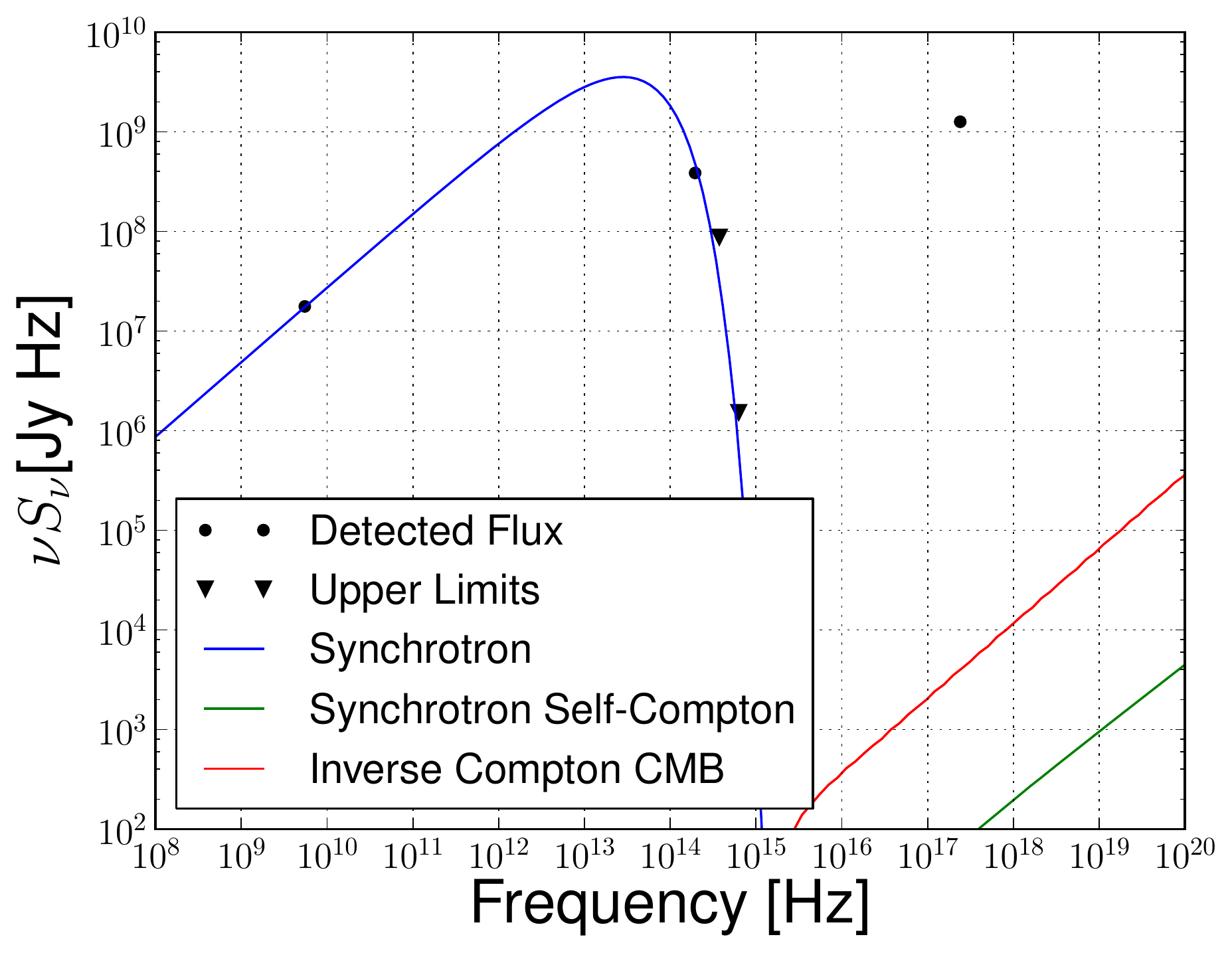}
		\caption{\label{fig:galaxysubtractions:SEDS_p150} $p=1.50$ }
	\end{subfigure} \\
	\begin{subfigure}{.49\linewidth}
		\includegraphics[width=\textwidth]{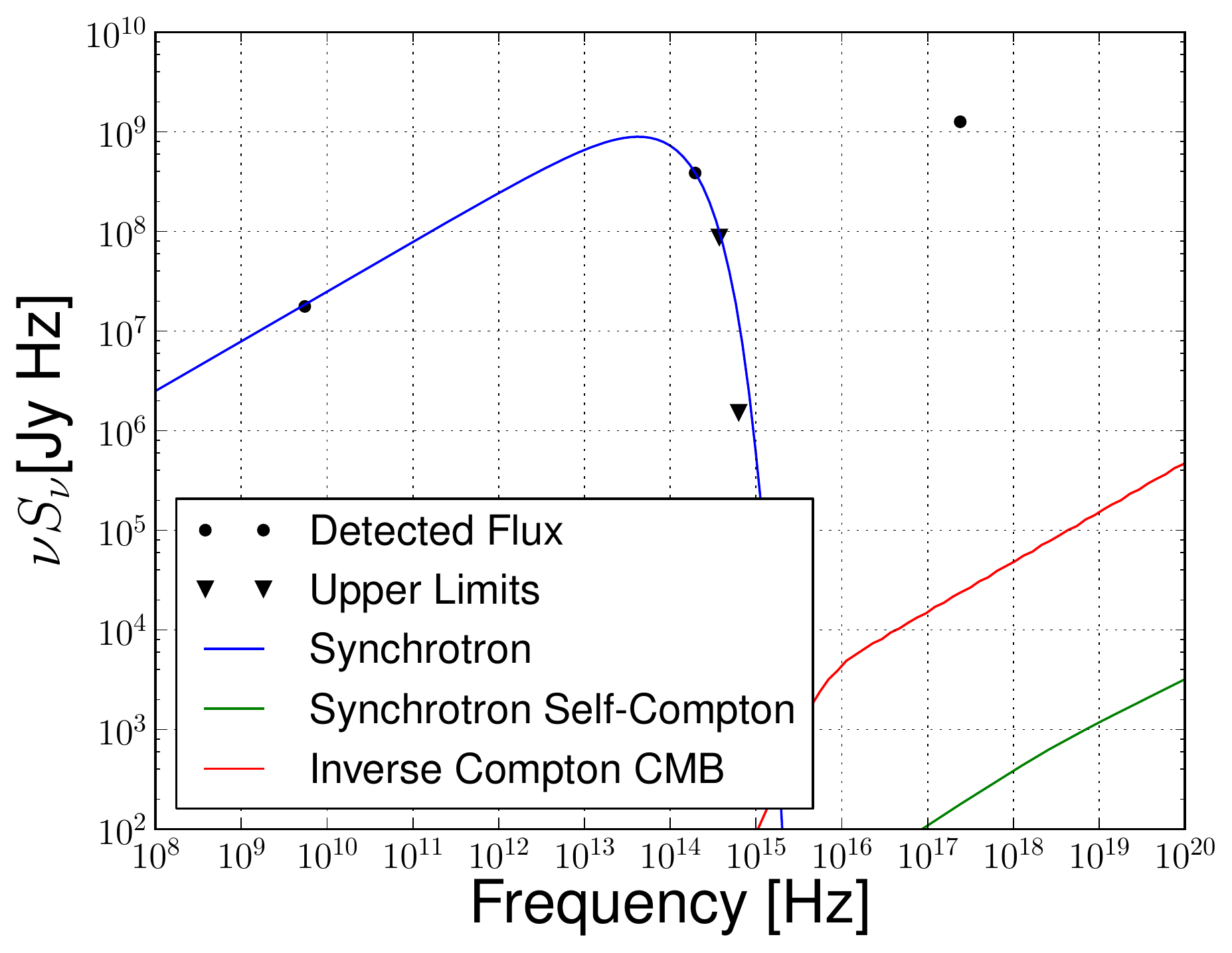}
		\caption{\label{fig:galaxysubtractions:SEDS_p200} $p=2.00$ }
	\end{subfigure}
	\begin{subfigure}{.49\linewidth}
		\includegraphics[width=\textwidth]{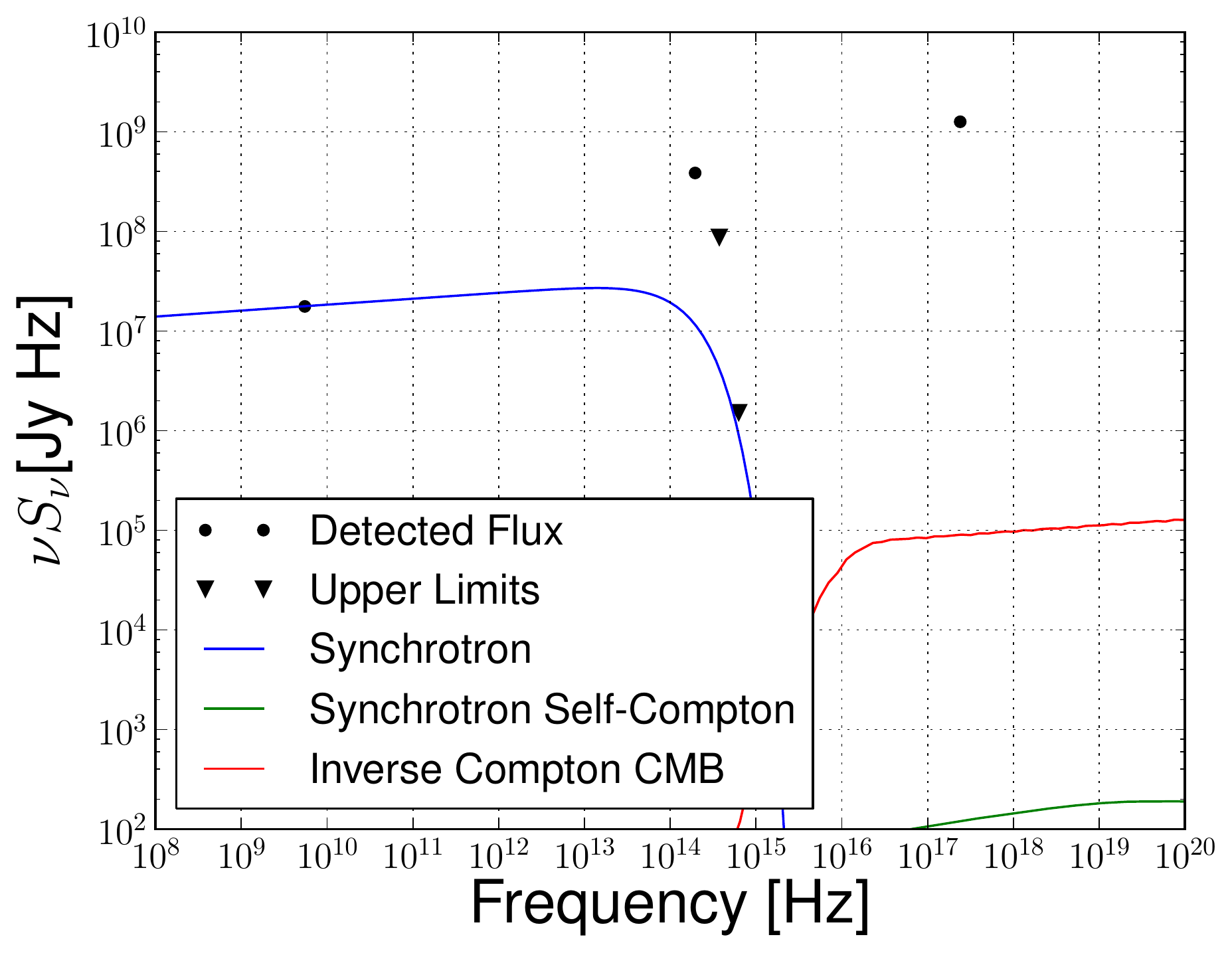}
		\caption{\label{fig:galaxysubtractions:SEDS_p288} $p=2.88$ suggested by \cite{Wilson01} }
	\end{subfigure}\\
	\caption{Computed SEDs for various electron indices, $p$. Notice that for $p\gtrsim 2.5$ (Figure \ref{fig:galaxysubtractions:SEDS_p288}; the index suggested by the results of \cite{Wilson01,Hardcastle05}), the flux doesn't rise enough after the radio emission to explain the observed F160W flux (without appealing to synchrotron self-absorption, which the data indicate is unlikely). \\
	Furthermore for values $p \gtrsim 2$, the synchrotron emission does not drop off quickly enough to explain an F160W detection with an F475W non-detection (\ref{fig:galaxysubtractions:SEDS_p200}). \\
	These models also support the prediction that IC-CMB emission (red) would dominate over SSC emission (green). Neither are sufficient though for explaining the x-ray emission, without Doppler boosting or other mechanisms.} 
	\label{fig:SEDS_p100_p150_p200_p288}
\end{figure}

Using the physical parameters in Table \ref{table:estimatedvalues}, we were able to numerically produce an SED for each electron index, $p$ (see Figure \ref{fig:SEDS_p100_p150_p200_p288}). The first result, which was predicted analytically, is that we need $p \lesssim 2.5$, in order to explain the detection of any knot flux in the F160W image, without requiring the radio flux to be in the self-absorbed regime. This constraint requires that the spectral index for the knot differ significantly from the x-ray indices for the entire jet found by \cite{Hardcastle05,Wilson01}.  Further strengthening this limit, we found $p \lesssim 2 $ in order to produce a sharp enough cutoff that could explain a detection in the F160W image, but not in the F814W or F475W images.

Finally, we briefly note that IC-CMB emission always dominates SSC emission, for all choices of $p$ at the x-ray frequency for which we have observations.  Neither IC-CMB nor SSC are sufficient for explaining the observed x-ray emission, so for that we must appeal to bulk Doppler boosting, or a different model entirely.

\subsection{Doppler Boosting}
\label{results:seds:DopplerBoosting}
The discrepancy between the predicted IC-CMB models and the observed x-ray flux suggests the presence of relativistic Doppler boosting.  We can use Equations \ref{eq:dopplerfactor} through \ref{eq:MarshallKbetamu}, but we must remember that we cannot uniquely invert Equation \ref{eq:MarshallKbetamu} if both $\Gamma$ and $\theta$ are unknowns.  By assuming one, we can solve for the other; by iterating through a number of assumed values, we can determine the extremal values which still predict real values. The results are found in Table \ref{table:dopplerboosting}, for values of $p$ which could explain the radio and optical data ($p \lesssim 2$).

\begin{table}[tbp]\centering
\ra{1.3}
\begin{tabular}{lllrrrr}\toprule
$p$  & \phantom{abc} & $\alpha = \frac{p-1}{2}$ & \phantom{abc} & $\Gamma_{min}$ & \phantom{abc} & $\theta_{max}$ \\  

\midrule

1.0  	&& 0.00 	&& 233    	&& 0.1		\\ 
1.5  	&& 0.25 	&& 41   	&& 0.6 		\\ 
2.001	&& 0.5005	&& 13   	&& 2.1  	\\

\bottomrule
\end{tabular}
\caption{\label{table:dopplerboosting} Parameters required for Doppler boosting to successfully describe x-ray emission via IC-CMB scattering. $\Gamma$ and $\theta$ are degenerate parameters in our observations, so we can only place limits, by finding the extremal values which yield real values for the other (see Equation \ref{eq:MarshallKbetamu}).  Furthermore, note that the inversion does not require $\Gamma_{min}$ to correspond to $\theta_{max}$.}
\end{table}

While the limits in Table \ref{table:dopplerboosting} allow for real values of $\Gamma$ and $\theta$, the systems they predict are unlikely and are not in-line with previous predications about this jet \cite{HSTProposal}.  In particular, with $\theta_{max}$ on the order of 1\arcdeg, we would expect the jet, roughly 1 kpc wide, to extend over 5 Mpc from the center of Pictor A, with beaming factors significantly larger than those found in a previous Chandra jet survey \cite{Marshall2005}.

In short, these optical images provide hard constraints on the electron index, $p$, for a successful synchrotron model for the primary knot.  Unfortunately, the parameters which could successfully explain the low frequency synchrotron radiation struggle to explain the significant x-ray flux.  This tension can be partially alleviated by introducing Doppler boosting, but it is unlikely that Doppler boosting can explain a majority of the high energy flux.

\chapter{Conclusions \& Significance}
\label{significance}

\section{Summarized Status of Radiative Transfer Hypotheses for Pictor A's Jet}
\label{significance:summary}
Early in the study of the jet of Pictor A, data strongly suggested that most of the observed flux was produced by non-thermal processes, such as synchrotron emission and inverse Compton scattering \cite{Wilson01}.  The precise nature of this emission was unknown though -- even by the time of Wilson's x-ray data, the results were not strong enough to identify the emission mechanism; in particular, he noted that he could not even rule out an unbroken spectral power law connecting radio and x-ray fluxes.

We have extended the understanding of Pictor A by obtaining optical data and using it to place meaningful constraints on the emission models of Pictor A's jet.  To start, our data conclusively rule out an unbroken power law, given the non-detections in the F814W and F475W images (see Figure \ref{fig:SED_nomodels_only32arcsec}); we need different emission mechanisms for different spectral regimes.  Spatially, we have shown that the parameters inferred from data of the entire jet differ significantly from parameters inferred from the primary knot, located 32\arcsec{} from the AGN.  This suggests we need a mechanism that explains distinguishably different electron populations at different locations along the jet.  Our results suggest that the electron population at the knot 32\arcsec{} from the AGN core is younger and has a harder spectrum than what we infer for the rest of the jet.  Such observations agree with a model of localized electron re-acceleration events (possibly due to shocks, the details of which are poorly understood).

Our data go on to disfavor the canonical model of a single population of electrons emitting non-thermally through synchrotron and inverse Compton processes.  We can successfully produce synchrotron models that describe the observed radio and optical data, but none of those models predict enough x-ray flux to match previous observations, even when reasonable levels of Doppler boosting are taken into account. 

We cannot definitively determine what is the source of the observed x-ray flux, but we can suggest a direction.  Hardcastle originally hypothesized that synchrotron radiation from a second population of electrons could explain the high levels of x-ray flux \cite{Hardcastle05}, a hypothesis motivated by similar results from a different FRII galaxy \cite{Kraft2005}. With the data that were available to Hardcastle, he could not rule out IC-CMB scattering as the primary x-ray producing mechanism, but that appears to have changed with our data.  With the addition of our optical data, we have shown it is unlikely for IC-CMB emission to be able to explain the observed x-ray flux. To continue the story of the study of Pictor A,, we suggest specific research in Appendix \ref{appendix:future} which would prove useful for our understanding Pictor A, and by extension, the study of extragalactic jets.

\appendix

\chapter{Knot Candidate Images}
\label{appendix:knotcandidateimages}

This appendix contains images of every knot candidate, in all optical bands for which we have data (F160W, F814W, F475W).  These images differ from the rest of the images in this work, as they have a \emph{linear} stretch function applied, rather than a \emph{logarithmic} stretch function (the much smaller field-of-view requires a much smaller dynamic range). The scales of these images do not necessarily correspond; these images are primarily to demonstrate the shape and the level of agreement between the data contours, in red, and the contours of the fitted 2d-gaussian models, in green (the contours for both are fixed to identical levels).  For all of the images, the jet runs from left to right.

\begin{figure}[p]
\centering
	\begin{subfigure}{.4\linewidth}
		\includegraphics[width=\textwidth]{f160w_32arcsec-eps-converted-to.pdf}
		\caption{F160W}
		\label{fig:knot_32arcsec_f160w} 
	\end{subfigure} \\
	\begin{subfigure}{.4\linewidth}
		\includegraphics[width=\textwidth]{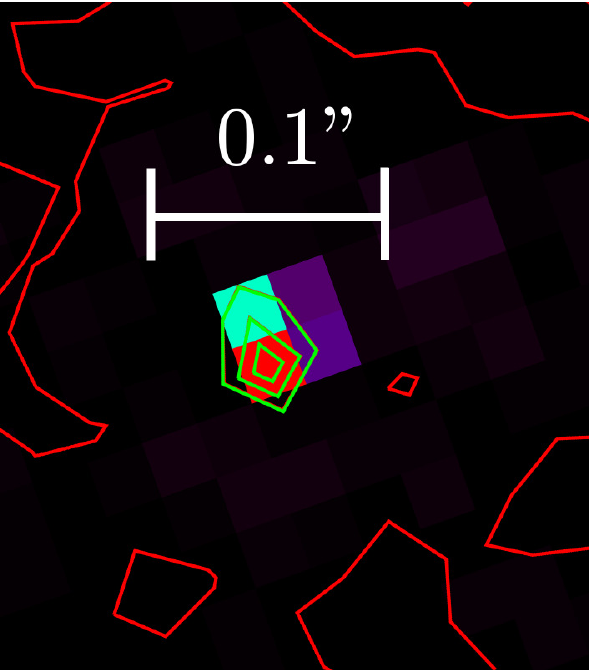}
		\caption{F814W -- Possible point source}
		\label{fig:knot_32arcsec_f814w}
	\end{subfigure} \\
	\begin{subfigure}{.4\linewidth}
		\includegraphics[width=\textwidth]{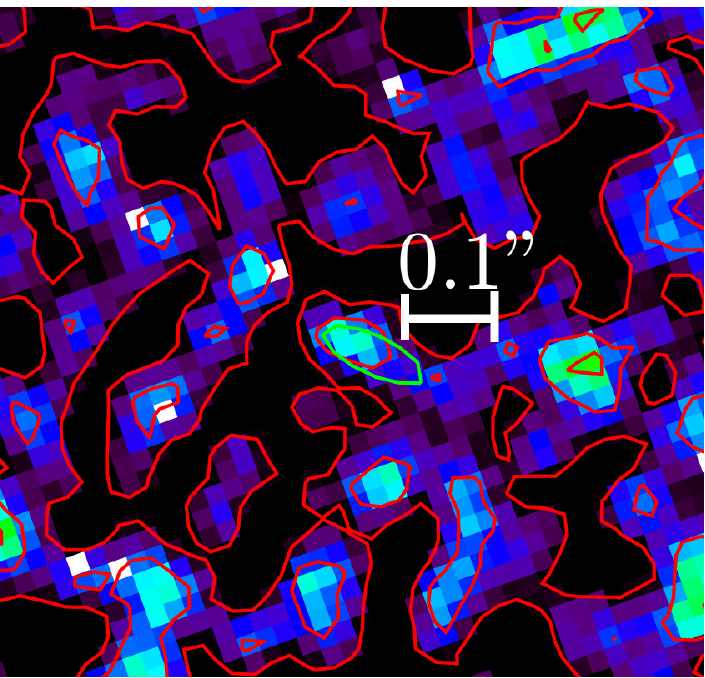}
		\caption{F475W -- Non-detection}
		\label{fig:knot_32arcsec_f475w}
	\end{subfigure}

	\caption{Knot located at 32\arcsec}
	\label{fig:knot_32arcsec}
\end{figure}

\begin{figure}[p]
\centering

	\begin{subfigure}{.4\linewidth}
		\includegraphics[width=\textwidth]{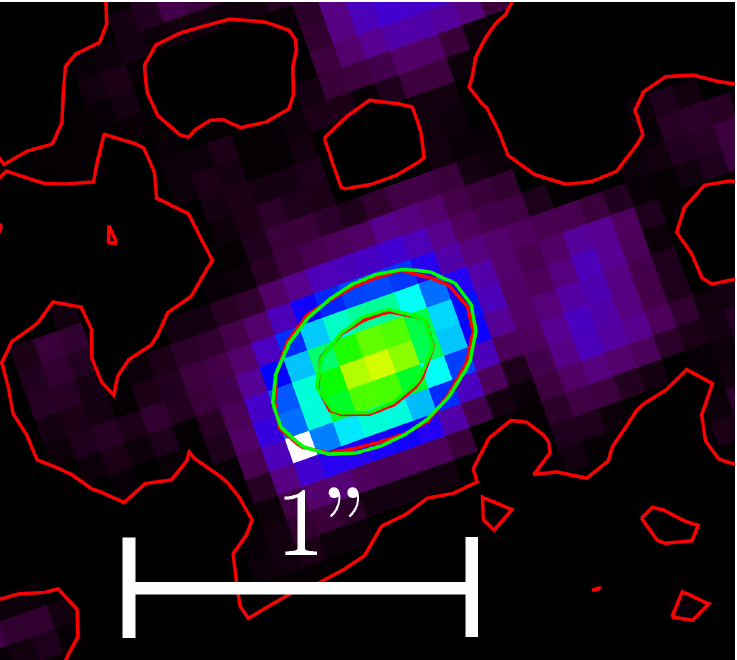}
		\caption{F160W}
		\label{fig:knot_43arcsec_f160w}
	\end{subfigure} \\
	\begin{subfigure}{.4\linewidth}
		\includegraphics[width=\textwidth]{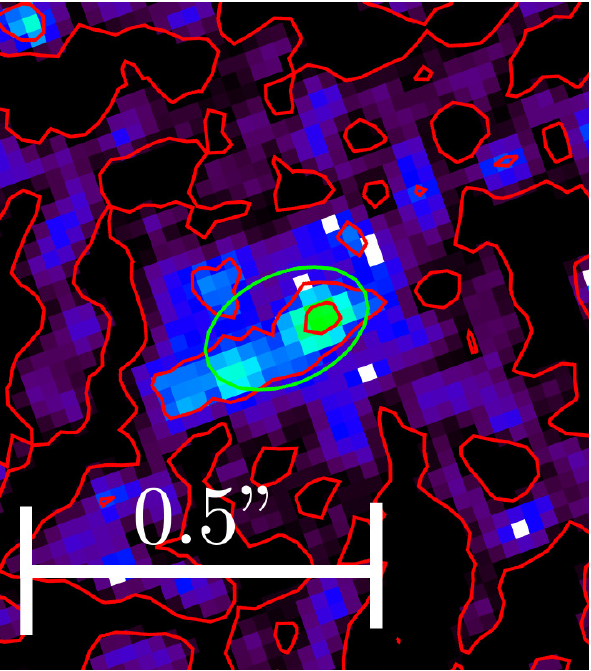}
		\caption{F814W}
		\label{fig:knot_43arcsec_f814w}
	\end{subfigure} \\
	\begin{subfigure}{.4\linewidth}
		\includegraphics[width=\textwidth]{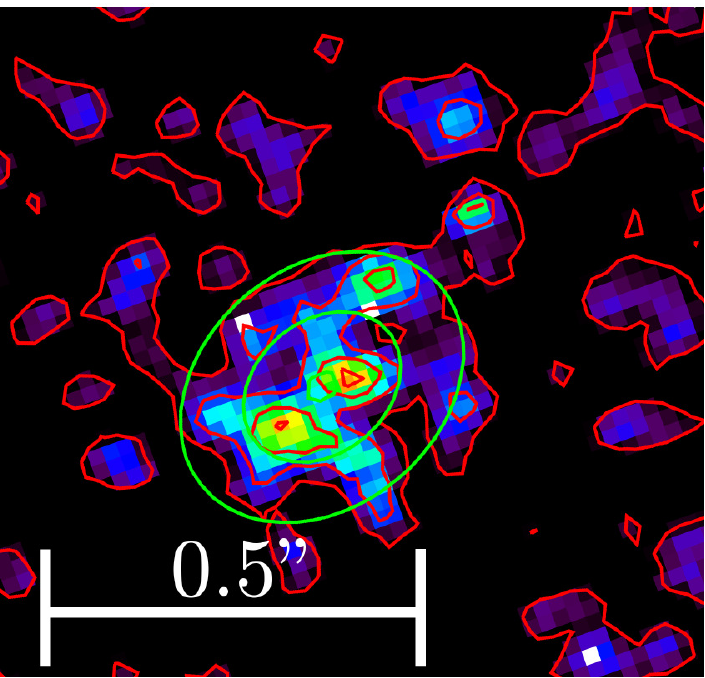}
		\caption{F475W}
		\label{fig:knot_43arcsec_f475w}
	\end{subfigure}

	\caption{Knot located at 43\arcsec}
	\label{fig:knot_43arcsec}
\end{figure}

\begin{figure}[p]
\centering

	\begin{subfigure}{.4\linewidth}
		\includegraphics[width=\textwidth]{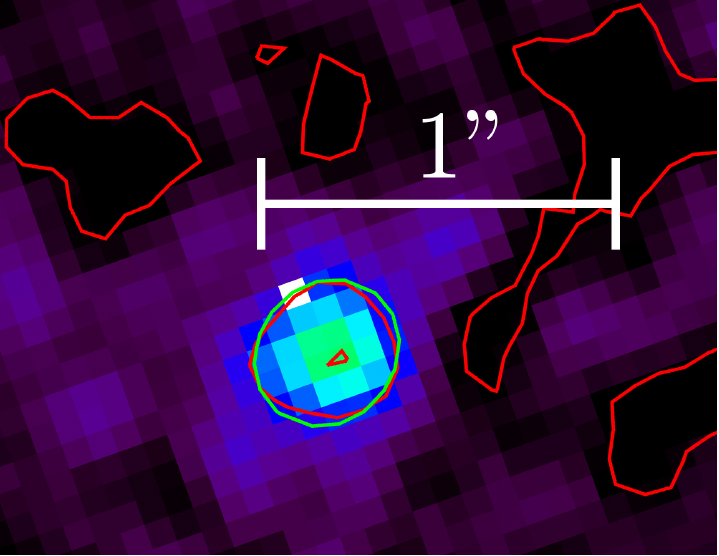}
		\caption{F160W}
		\label{fig:knot_106arcsec_f160w}
	\end{subfigure} \\
	\begin{subfigure}{.4\linewidth}
		\includegraphics[width=\textwidth]{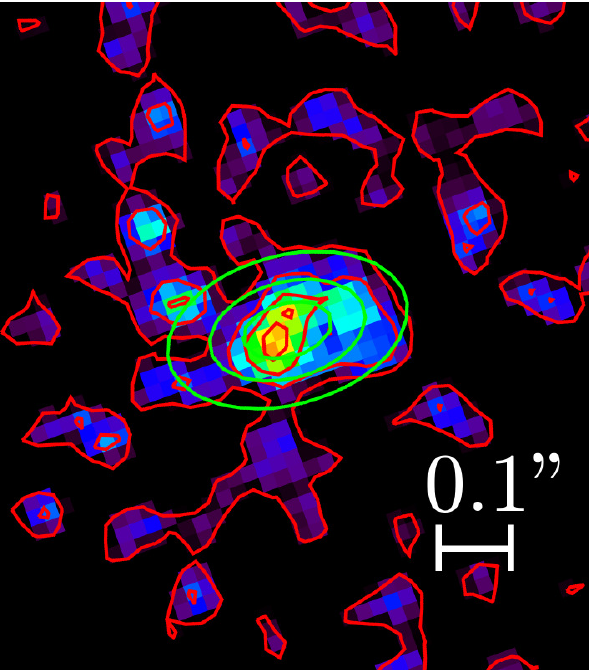}
		\caption{F814W}
		\label{fig:knot_106arcsec_f814w}
	\end{subfigure} \\
	\begin{subfigure}{.4\linewidth}
		\includegraphics[width=\textwidth]{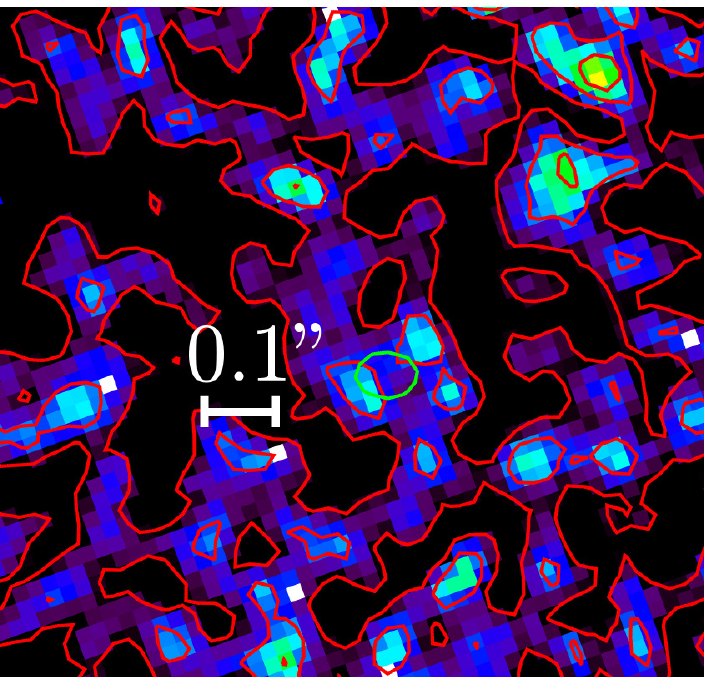}
		\caption{F475W}
		\label{fig:knot_106arcsec_f475w}
	\end{subfigure}

	\caption{Knot located at 106\arcsec}
	\label{fig:knot_106arcsec}
\end{figure}

\begin{figure}[p]
\centering

	\begin{subfigure}{.4\linewidth}
		\includegraphics[width=\textwidth]{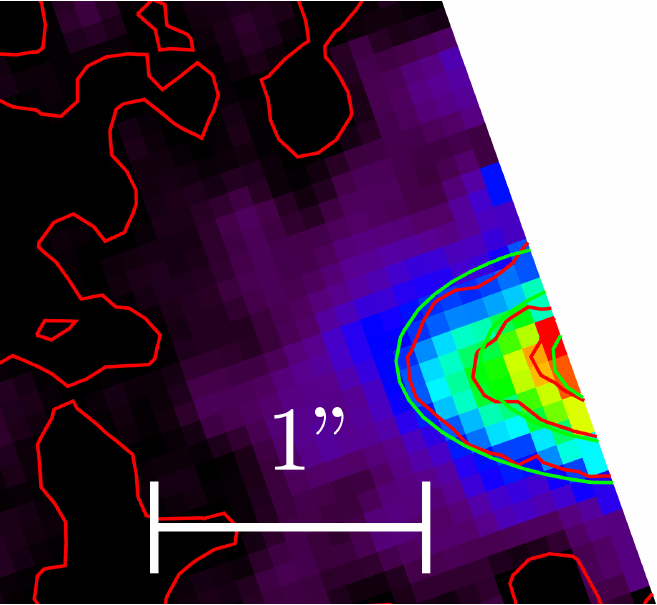}
		\caption{F160W -- Diagonal cutoff is from the edge of the detector}
		\label{fig:knot_112arcsec_f160w}
	\end{subfigure} \\
	\begin{subfigure}{.4\linewidth}
		\includegraphics[width=\textwidth]{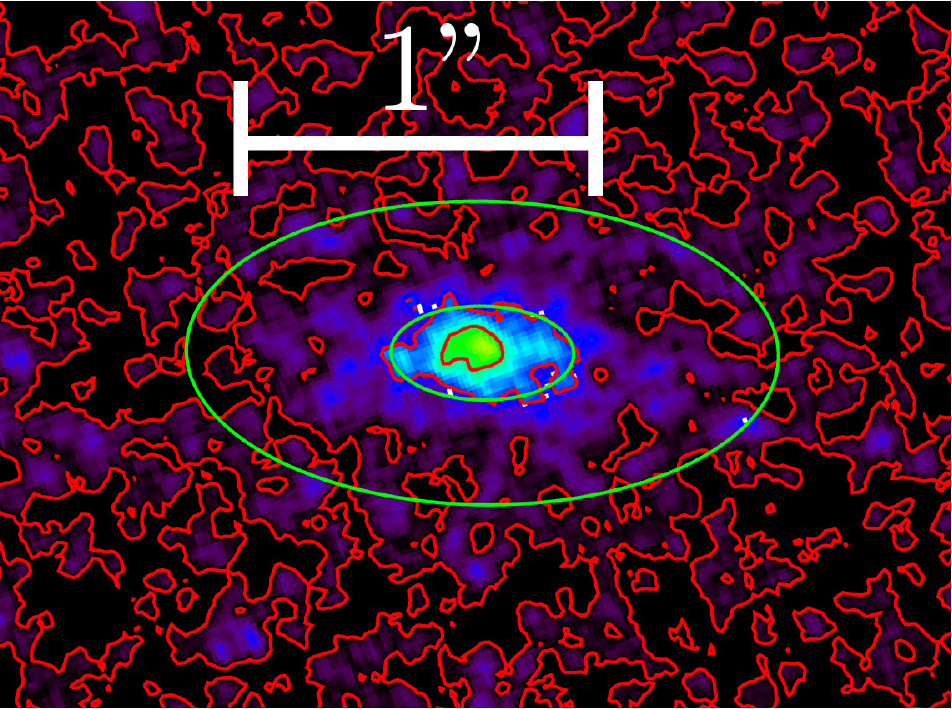}
		\caption{F814W}
		\label{fig:knot_112arcsec_f814w}
	\end{subfigure} \\
	\begin{subfigure}{.4\linewidth}
		\includegraphics[width=\textwidth]{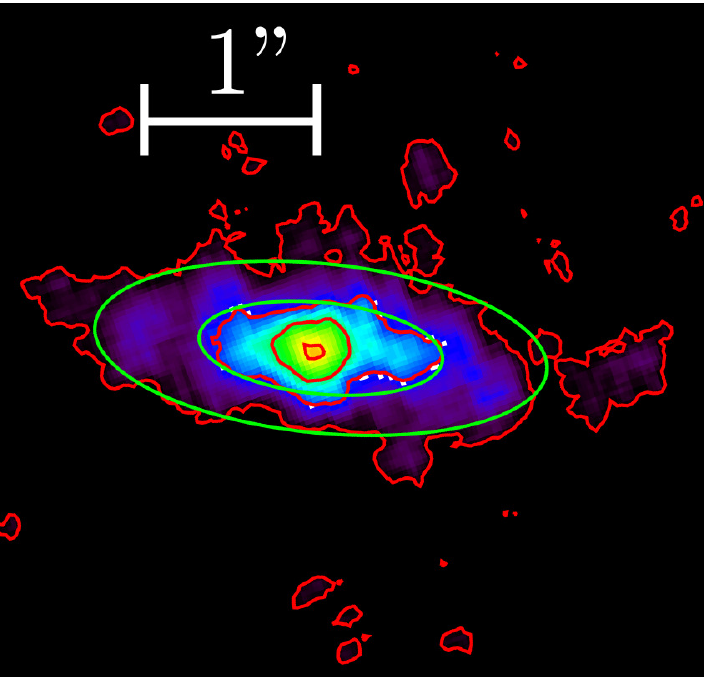}
		\caption{F475W}
		\label{fig:knot_112arcsec_f475w}
	\end{subfigure}
	\caption{Knot located at 112\arcsec}
	\label{fig:knot_112arcsec}
\end{figure}

\clearpage
\newpage


\chapter{Future Work}
\label{appendix:future}
The study of Pictor A continues to confound the typical models used to explain AGN and jet emission. This suggests there is something to be learned, something beyond the most popular models of the physical processes within jets.  While further analyses will continue in preparation for a journal publication of these results, there are key questions which go beyond the scope of this work.  In order to better understand the perplexing Pictor A, we suggest future analysis which could be done on existing data (Section \ref{future:tidaltail}) and new observations which could add significant data (Section \ref{future:observations}). 

\section{Tidal Tail Analysis}
\label{future:tidaltail}
These observations discovered a previously unknown and unexpected tidal tail (this tidal tail is the broad, sweeping feature North of Pictor A, most clearly visible in the F160W image; see Figure \ref{fig:galaxysubtractions_largefont}).  This feature suggests a previous merger event in the history of Pictor A, and it might provide a history of the AGN and jet which we are studying. One hypothesis is that galaxy merger could play a role in triggering active galaxies; while this hypothesis does not appear to explain most AGN \cite{Villforth2014,Cisternas2011}, Pictor A could provide information about those AGN for which a merger might have been significant.

\section{Suggested Observations}
\label{future:observations}
As noted in Section \ref{significance:summary}, our ability to constrain emission models of Pictor A's jet remains largely limited by the available observational data. We suggest future observations which could break some of the degeneracies left by the current data.

\subsection{New Spectral Bands}
\label{future:observations:newspectralbands}
While observations stretch from radio to optical to x-ray bands, we are still missing crucial information for constraining synchrotron parameters.  In particular, it would be useful to know the spectral index at radio frequencies, since such a determination could provide significant evidence for 2 distinct electron populations, if radio and x-ray spectral indices disagree.

Our synchrotron models predict radio emission to become brighter with increasing frequency, leading us to suggest the Atacama Large Millimeter/sub-millimeter Array (\emph{ALMA}).  ALMA has recently begun operating with spectral bands as high as 900 GHz, compared to our current radio map taken at about 5 GHz. Furthermore ALMA is well-positioned at a latitude of 23\arcdeg{} South for observing Pictor A, which lies at a declination of $45\arcdeg$ South (outside the range of most Northern Hemisphere observatories).  ALMA could leverage the brighter emission at higher frequencies to break key degeneracies in our spectral models of Pictor A's jet.

\subsection{Deeper Observations}
\label{future:observations:deeperobservations}
\begin{figure}[tbp]
	\begin{center}
		\includegraphics[width=.8\linewidth]{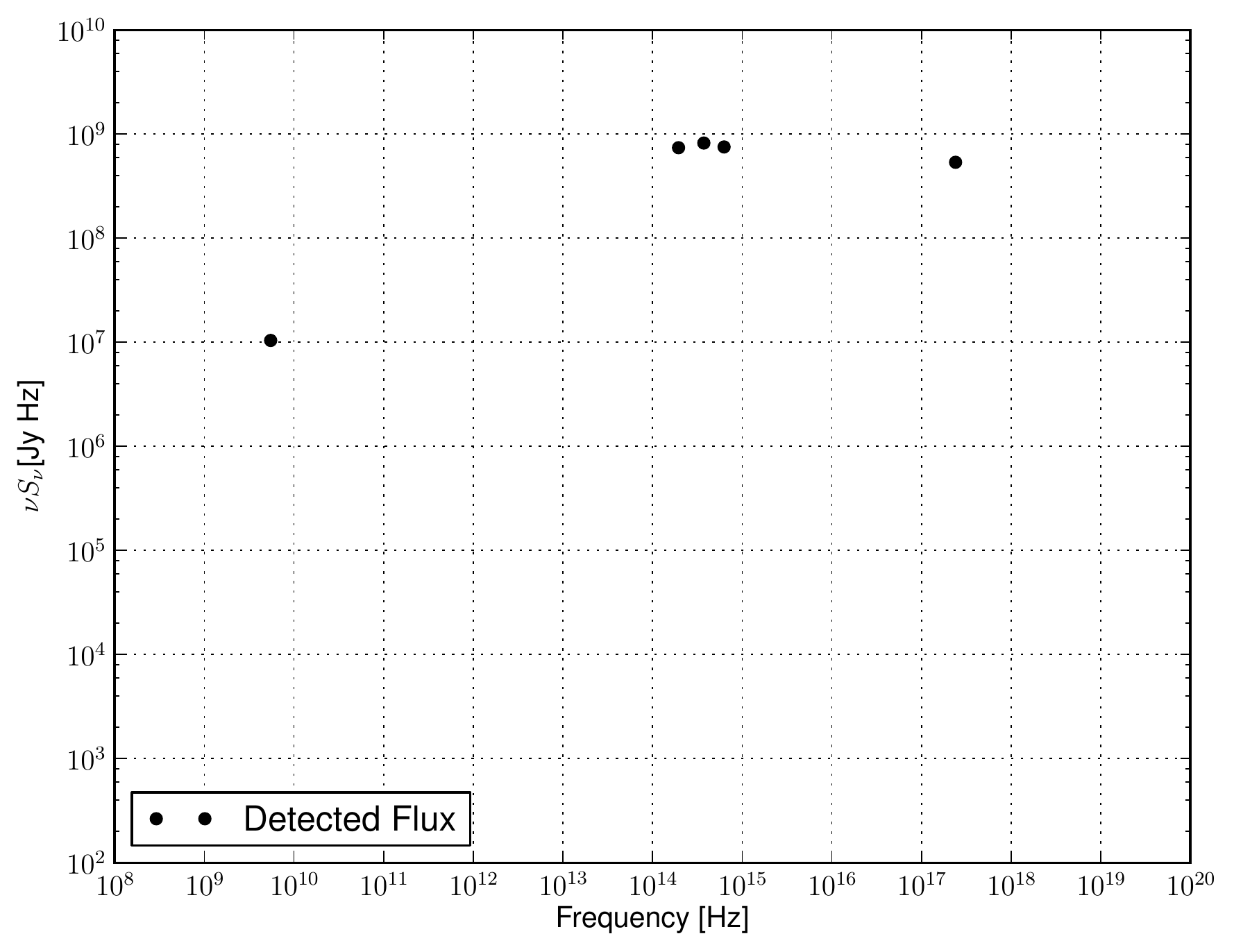}
	\end{center}
	\caption{SED of data only from the 112\arcsec{} knot. The F160W data in particular should be treated cautiously -- by trying to fit a 2d-gaussian, with only half of the profile visible (see Figure \ref{fig:knot_112arcsec_f160w}) we introduce potentially large, uncharacterized systematic errors. }
	\label{fig:SED_nomodels_112arcsec}
\end{figure}

Deeper HST observations, while not breaking the degeneracy of radio spectral indices, might generate a greater number of knot detections.  Many of the observed knot candidates (particularly those at 43\arcsec{} and 106\arcsec) were only marginally detected.  Furthermore, the knot candidate at 112\arcsec{} appears to have a high flux (even in the existing F475W image), but in the current images, this knot lies partially beyond the edge of the F160W detector.  Using the information we have gained, the pointing of the HST could easily be adjusted to capture that knot fully, on every detector.  By fitting what we can see of the knot in the F160W image, it appears that it has a higher synchrotron cutoff than the other observed knots (compare the 112\arcsec{} SED in Figure \ref{fig:SED_nomodels_112arcsec} with the SED for the other observed knots in Figure \ref{fig:SED_nomodels_allknots}; see the images of the 112\arcsec{} knot in Figure \ref{fig:knot_112arcsec}).  Not only does this mean the knot 112\arcsec{} from the core could yield different information about the jet, but it also means we are very likely to make a strong detection in all bands, given the high flux being produced.

\begin{singlespace}
\bibliography{main}
\bibliographystyle{plain}
\end{singlespace}

\end{document}